\documentclass[multphys,vecphys]{svmult}

\usepackage{makeidx}     % allows index generation
\usepackage{graphicx}    % standard LaTeX graphics tool
\usepackage{multicol}    % used for the two-column index
\makeindex             % used for the subject index
\def\simgt{\lower 2pt \hbox{$\, \buildrel {\scriptstyle >}\over {\scriptstyle \sim}\,$}}
\def\simlt{\lower 2pt \hbox{$\, \buildrel {\scriptstyle <}\over {\scriptstyle \sim}\,$}}
\def\asca{{\it ASCA\/}}
\def\chandra{{\it Chandra\/}}
\def\conx{{\it Constellation-X\/}}
\def\einstein{{\it Einstein\/}}
\def\genx{{\it Generation-X\/}}
\def\heao1{{\it {\it HEAO1}\/}}
\def\hst{{\it {\it HST}\/}}

\def\rosat{{\it ROSAT\/}}
\def\sax{{\it BeppoSAX\/}}
\def\spitzer{{\it Spitzer\/}}
\def\swift{{\it Swift\/}}
\def\xeus{{\it XEUS\/}}
\def\xmm{{\it XMM-Newton\/}}

\begin{document}

\title*{X-ray Survey Results on Active Galaxy Physics and Evolution}
\titlerunning{X-ray Survey Results on AGN Physics and Evolution}
\author{W.N. Brandt,\inst{1} D.M. Alexander,\inst{2} F.E. Bauer,\inst{2} \and C. Vignali\inst{3}}
\institute{Department of Astronomy \& Astrophysics, The Pennsylvania 
State University, 525 Davey Lab, University Park, PA 16802, USA
\texttt{(niel@astro.psu.edu)}
\and
Institute of Astronomy, Madingley Road, Cambridge CB3 0HA, UK
\texttt{(dma@ast.cam.ac.uk, feb@ast.cam.ac.uk)}
\and
INAF---Osservatorio Astronomico di Bologna, Via Ranzani 1, 
40127 Bologna, Italy
\texttt{(cristian@anastasia.bo.astro.it)}}
\maketitle

% --------------------------------------------------------------------------------

\section{Introduction}
\label{sec:1}

% Always give a unique label
% and use \ref{<label>} for cross-references
% and \cite{<label>} for bibliographic references
% use \sectionmark{}
% to alter or adjust the section heading in the running head
% Your text goes here. Use the \LaTeX\ automatism for your citations
% \cite{monograph}.

The cosmic \hbox{X-ray} background (XRB) is largely due to accretion onto
supermassive black holes integrated over cosmic time. Thus,  
extragalactic \hbox{X-ray} surveys offer the potential to contribute 
substantially to our understanding of the physics 
of Active Galactic Nuclei (AGN) as well as the 
evolution of the AGN population. Such surveys have dramatically advanced 
over the past four years, largely due to the flood of data from 
the {\it Chandra X-ray Observatory\/} (hereafter \chandra) and the 
{\it X-ray Multi-Mirror Mission-Newton\/} (hereafter \xmm). The superb \hbox{X-ray} 
mirrors and charge-coupled device (CCD) detectors on these observatories 
provide

\begin{description}

\item{$\bullet$}
Sensitive imaging spectroscopy in the \hbox{$\approx 0.5$--10~keV} band, 
with up to \hbox{50--250} times (depending upon the energy band considered) 
the sensitivity of previous \hbox{X-ray} missions. \hbox{X-ray} surveys have
finally reached the depths needed to complement the most
sensitive surveys in the radio, submillimeter, infrared, and
optical bands. 
   
\item{$\bullet$}
\hbox{X-ray} source positions with accuracies of 
\hbox{$\approx$~0.3--3$^{\prime\prime}$}. These high-quality positions 
are essential for matching to (often faint) multiwavelength counterparts. 

\item{$\bullet$}
Large source samples allowing reliable statistical inferences
to be drawn about the overall extragalactic \hbox{X-ray} source 
population. In a fairly deep \chandra\ or \xmm\ observation, 
$\simgt 100$--200 sources can be detected. 

\end{description}

\noindent
The extragalactic survey capabilities of \chandra\ and \xmm\ are 
complementary in several important respects. The sub-arcsecond imaging
of \chandra\ provides the best possible source positions, and with 
long exposures \chandra\ can achieve the highest possible sensitivity 
at energies of \hbox{$\approx$~0.5--6~keV}; unlike the case for \xmm, even 
the deepest \chandra\ observations performed to date do not suffer from 
significant source confusion. \xmm, in comparison, has a substantially
larger photon collecting area than \chandra, allowing efficient
\hbox{X-ray} spectroscopy. In addition, \xmm\ has better high-energy response
than \chandra\ and can carry out the deepest possible surveys from
\hbox{$\approx$~7--10~keV}. Even \xmm, however, does not cover the peak of 
the \hbox{X-ray} background at \hbox{20--40~keV}. Finally, the
field of view for \xmm\ is $\sim 2.5$ times that of \chandra. 

\begin{table}[t!]
\centering
\caption{Some Deep X-ray Surveys with \chandra\ and \xmm}
\label{tab:1}       % Give a unique label
\begin{tabular}{lrl}
\hline\noalign{\smallskip}
Survey Name & Exposure & Representative Reference or Note  \\
\noalign{\smallskip}\hline\noalign{\smallskip}
\noalign{\chandra}
\noalign{\smallskip}\hline\noalign{\smallskip}
\chandra\ Deep Field-North          & 1950~ks \hspace{0.1 in} & D.M. Alexander et~al., 2003, AJ, 126, 539 \\          
\chandra\ Deep Field-South          &  940~ks \hspace{0.1 in} & R. Giacconi et~al., 2002, ApJS, 139, 369 \\
HRC Lockman Hole                    &  300~ks \hspace{0.1 in} & PI: Murray \\
Extended CDF-S                      &  250~ks \hspace{0.1 in} & PI: Brandt \\
Groth-Westphal                      &  200~ks \hspace{0.1 in} & PI: Nandra \\
Lynx                                &  185~ks \hspace{0.1 in} & D. Stern et~al., 2002, AJ, 123, 2223 \\           
LALA Cetus                          &  177~ks \hspace{0.1 in} & PI: Malhotra \\
LALA Bo\"otes                       &  172~ks \hspace{0.1 in} & J.X. Wang et~al., 2004, AJ, 127, 213 \\
SSA13                               &  101~ks \hspace{0.1 in} & A.J. Barger et~al., 2001, AJ, 121, 662 \\
3C295                               &  100~ks \hspace{0.1 in} & V. D'Elia et~al., 2004, astro-ph/0403401 \\
Abell~370                           &   94~ks \hspace{0.1 in} & A.J. Barger et~al., 2001, AJ, 122, 2177 \\
SSA22 ``protocluster''              &   78~ks \hspace{0.1 in} & L.L. Cowie et~al., 2002, ApJ, 566, L5 \\
ELAIS                               &   75~ks \hspace{0.1 in} & J.C. Manners et~al., 2003, MNRAS, 343, 293  \\
WHDF                                &   75~ks \hspace{0.1 in} & PI: Shanks \\
\noalign{\smallskip}\hline\noalign{\smallskip}
\noalign{\xmm}
\noalign{\smallskip}\hline\noalign{\smallskip}
Lockman Hole                  &  766~ks \hspace{0.1 in} &  G. Hasinger et~al., 2001, A\&A, 365, L45 \\
\chandra\ Deep Field-South    &  317~ks \hspace{0.1 in} &  A. Streblyanska et~al., 2004, astro-ph/0309089 \\
13~hr Field                   &  200~ks \hspace{0.1 in} &  M.J. Page et~al., 2003, AN, 324, 101 \\
\chandra\ Deep Field-North    &  180~ks \hspace{0.1 in} &  T. Miyaji et~al., 2003, AN, 324, 24 \\
Subaru Deep                   &  100~ks \hspace{0.1 in} &  PI: Watson \\
ELAIS S1                      &  100~ks \hspace{0.1 in} &  PI: Fiore \\
Groth-Westphal                &   80~ks \hspace{0.1 in} &  T. Miyaji et~al., 2004, astro-ph/0402617 \\
\noalign{\smallskip}\hline
\end{tabular}
\\
\raggedright
{The Extended \chandra\ Deep Field-South is comprised of four fields (each 250~ks), 
the \xmm\ ELAIS S1 survey is comprised of four fields (each 100~ks), and
the \chandra\ ELAIS survey is comprised of two fields (each 75~ks). The \xmm\ 
Subaru Deep survey also has seven flanking fields (each $\approx 50$~ks).
Only the first $\approx 100$~ks of the \xmm\ Lockman Hole data have been
published at present.}
\end{table}

\chandra\ and \xmm\ have resolved $\simgt 80$--90\%\ of the \hbox{0.5--10~keV} 
XRB into discrete sources, extending earlier heroic efforts with 
missions including \rosat, \asca, and \sax. The main uncertainties in
the precise resolved fraction are due to 
field-to-field cosmic variance (which leads to spatial variation in the
XRB flux density) and instrumental cross-calibration limitations. 
With the recent advances, attention is now 
focused on (1) understanding the nature of the \hbox{X-ray} sources
in detail and their implications for AGN physics, and (2) understanding
the cosmological evolution of the sources and their role in galaxy 
evolution. In this review, we briefly describe the key \chandra\ and \xmm\ 
extragalactic surveys to date (\S2) and detail some of their implications 
for AGN physics and evolution (\S3). In \S3 we highlight two topics
of current widespread interest: 
(1) X-ray constraints on the AGN content of luminous submillimeter 
galaxies, and
(2) the demography and physics of high-redshift $(z>4)$ AGN as
revealed by X-ray observations. We also discuss prospects for future 
\hbox{X-ray} surveys with \chandra, \xmm, and upcoming missions (\S4). 

Throughout this paper, we adopt
$H_0=70$~km~s$^{-1}$ Mpc$^{-1}$, 
$\Omega_{\rm M}=0.3$, and 
$\Omega_{\Lambda}=0.7$ (flat cosmology). 

\begin{figure}[t!]
\centering
\includegraphics[height=10cm]{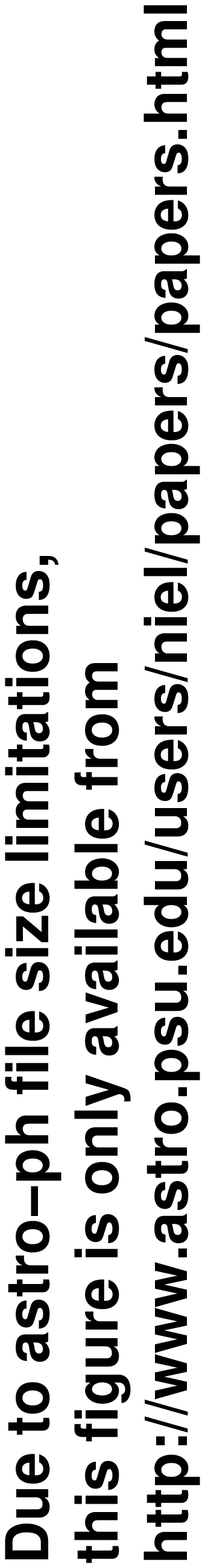}
\caption{Adaptively smoothed image of the 2~Ms \hbox{CDF-N}, constructed from
data in the \hbox{0.5--2~keV} (red), \hbox{2--4~keV} (green), and \hbox{4--8~keV} 
(blue) bands. Nearly 600 sources are detected in the $\approx 448$~arcmin$^2$ field. 
The regions covered by the HDF-N and GOODS-N surveys are denoted. Adapted 
from D.M. Alexander, F.E. Bauer, W.N. Brandt, et~al., 2003, AJ, 126, 539.}
\label{fig:1}       
\end{figure}

% --------------------------------------------------------------------------------

\section{Chandra and XMM-Newton Extragalactic Surveys}
\label{sec:2}

To learn about AGN physics and evolution in a complete manner, both
``deep'' and ``wider'' \hbox{X-ray} surveys are required; the trade-off between
the two, of course, is between sensitivity and solid-angle coverage on the 
sky. None of the \chandra\ and \xmm\ surveys discussed in this paper is 
truly wide-field, in that the widest still only cover $\simlt 1$\%\ of 
the sky.\footnote{For this reason, we denote these surveys as ``wider''
(relative to the deep \chandra\ and \xmm\ surveys) rather than ``wide-field.''} 
Both deep and wider \hbox{X-ray} surveys are reviewed briefly below. 

% --------------------------------------------------------------------------------

\subsection{Deep X-ray Surveys}
\label{sec:2p1}

Table~1 makes it clear that deep \chandra\ and \xmm\ surveys are a
major ``industry.'' The 21 surveys listed there have a total exposure
exceeding 70~days, and $\simgt 50$ scientists have invested substantial
effort on the analysis and interpretation of these data. Comparable
effort has also been expended on multiwavelength follow-up studies 
of these surveys; due to the small solid angles under investigation, superb 
multiwavelength coverage can be obtained relatively economically. 

\begin{table}[t!]
\centering
\caption{Some Wider X-ray Surveys with \chandra\ and \xmm}
\label{tab:2}       % Give a unique label
\begin{tabular}{lll}
\hline\noalign{\smallskip}
Survey Name & $\Omega$ (deg$^2$) \hspace{0.1 in} & Representative Reference or Note  \\
\noalign{\smallskip}\hline\noalign{\smallskip}
\noalign{\chandra}
\noalign{\smallskip}\hline\noalign{\smallskip}
ChaMP                       \hspace{0.1 in} & 14       &    D.W. Kim et~al., 2004, ApJS, 150, 19 \\
Clusters Serendipitous      \hspace{0.1 in} & 1.1      &    P. Gandhi et~al., 2004, MNRAS, 348, 529 \\
CYDER                       \hspace{0.1 in} & $\cdots$ &    F.J. Castander et~al., 2003, AN, 324, 40 \\
Lockman Hole NW             \hspace{0.1 in} & 0.4      &    A.T. Steffen et~al., 2003, ApJ, 596, L23 \\
MUSYC                       \hspace{0.1 in} & 1        &    PI: van Dokkum \\
NOAO DWFS                   \hspace{0.1 in} & 9.3      &    PI: Jones \\
SEXSI                       \hspace{0.1 in} & 2.2      &    F.A. Harrison et~al., 2003, ApJ, 596, 944 \\
SWIRE Lockman               \hspace{0.1 in} & 0.6      &    PI: Wilkes \\
1~hr Field                  \hspace{0.1 in} & 0.2      &    PI: McHardy \\
13~hr Field                 \hspace{0.1 in} & 0.2      &    I.M. McHardy et~al., 2003, MNRAS, 342, 802 \\
\noalign{\smallskip}\hline\noalign{\smallskip}
\noalign{\xmm}
\noalign{\smallskip}\hline\noalign{\smallskip}
AXIS                  \hspace{0.1 in} & $\cdots$  &    X. Barcons et~al., 2002, A\&A, 382, 522 \\
CFRS                  \hspace{0.1 in} & 0.6       &    T.J. Waskett et~al., 2003, MNRAS, 341, 1217 \\
HELLAS2XMM            \hspace{0.1 in} & 2.9       &    A. Baldi et~al., 2002, ApJ, 564, 190 \\
LSS                   \hspace{0.1 in} & 64        &    M. Pierre et~al., 2004, astro-ph/0305191 \\
Survey Science Center \hspace{0.1 in} & $\cdots$  &    M.G. Watson et~al., 2001, A\&A, 365, L51 \\
VIMOS                 \hspace{0.1 in} & 2.3       &    PI: Hasinger \\
2dF                   \hspace{0.1 in} & 1.5       &    A. Georgakakis et~al., 2003, MNRAS, 344, 161 \\
\noalign{\smallskip}\hline
\end{tabular}
\\
\raggedright
{The second column above lists estimated survey solid angles; 
survey sensitivities are not uniform but rather vary significantly
across these solid angles. In some cases, survey solid angles
are not well defined and thus are not listed. In these cases, the
reader should consult the listed reference or note for further
details.}
\end{table}

\begin{figure}[t!]
\centering
\includegraphics[height=9cm]{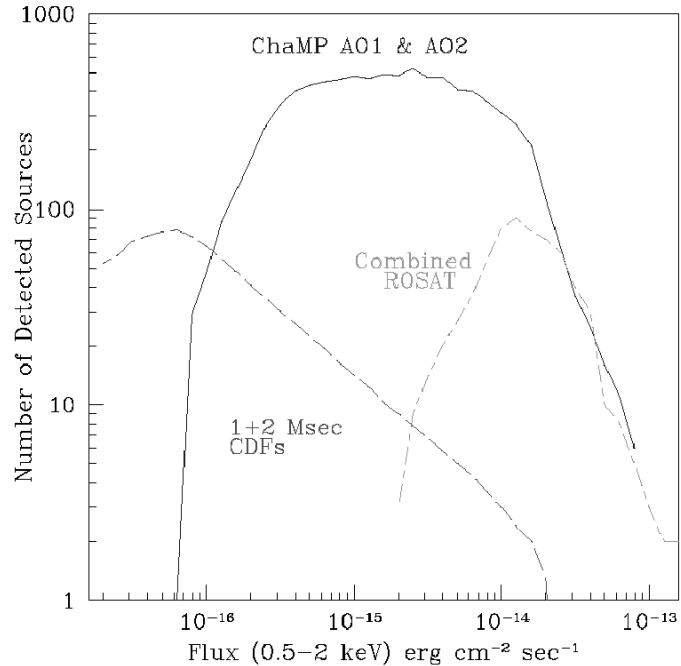}
\caption{Number of sources predicted from the ChaMP survey (for 137
ChaMP fields from \chandra\ Cycle~1 and Cycle~2) compared to numbers
of sources from the \chandra\ Deep Fields and the \rosat\ surveys
analyzed by T. Miyaji, G. Hasinger, \& M. Schmidt, 2000, A\&A, 353, 25. An 
impressive $\approx 6000$ ChaMP sources are expected 
in total, and these largely lie at intermediate \hbox{0.5--2~keV} flux 
levels of \hbox{(4--60)$\times 10^{-16}$~erg~cm$^{-2}$~s$^{-1}$}. 
From D.W. Kim, R.A. Cameron, J.J. Drake, et~al., 2004, ApJS, 150, 19.}
\label{fig:2}       
\end{figure}

The two most sensitive surveys in Table~1, by a 
significant factor, are the 2~Ms \chandra\ Deep Field-North (\hbox{CDF-N}; see 
Figure~1) and 1~Ms \chandra\ Deep Field-South (\hbox{CDF-S}). 
Both are situated in intensively studied regions of sky with little
Galactic foreground \hbox{X-ray} absorption. They reach \hbox{0.5--2~keV} 
fluxes of $\approx$~(2.5--5)$\times 10^{-17}$~erg~cm$^{-2}$~s$^{-1}$, 
corresponding to count rates of $\simlt 1$ count every \hbox{2--4} days. At
these flux levels, even moderate-luminosity AGN (similar to Seyfert
galaxies in the local universe) can be detected to $z\simgt 10$. 
The \hbox{CDF-N} and \hbox{CDF-S} are clearly ``pencil-beam'' surveys, each covering 
$\approx 400$~arcmin$^{2}$; for reference, this is $\sim 1/2$ the solid 
angle of the full Moon and $\sim 75$ times the solid angle of the original 
Hubble Deep Field-North (HDF-N; see Figure~1). Public \hbox{X-ray} 
source catalogs are available for both the \hbox{CDF-N} 
and \hbox{CDF-S} (see the references in Table~1); these contain 
$\approx 580$ and $\approx 370$ sources, respectively.  

The other deep \hbox{X-ray} surveys in Table~1 have generally been performed
in regions of sky where (1) extensive coverage already exists at
one-to-several wavelengths, and/or (2) some interesting astronomical
object is present (e.g., 3C295, Abell~370, or the SSA22 ``protocluster''). 
They are all sensitive enough to detect moderate-luminosity AGN to 
$z\sim 3$--5, and in total the surveys in Table~1 cover a solid angle
of $\sim 3.5$~deg$^2$ ($\sim 16$ Moons). 

% --------------------------------------------------------------------------------

\subsection{Wider X-ray Surveys}
\label{sec:2p2}

Wider \chandra\ and \xmm\ surveys (see Table~2) are a comparably large and
important ``industry'' to the deep surveys. These typically involve
investigation of \hbox{X-ray} sources in \hbox{$\sim 4$--150} \hbox{X-ray} observations of
moderate exposure \hbox{(usually 20--60~ks, but sometimes as short as $\approx 5$~ks)}; 
the observations are sometimes obtained from the public 
data archives. The wider surveys serve to bridge the observational ``gap''
between the deepest \chandra\ observations and the deepest observations
made by previous \hbox{X-ray} missions (e.g., \rosat; see Figure~2), and they 
effectively target the intermediate 0.5--8~keV flux levels 
($10^{-15}$--$10^{-13}$~erg~cm$^{-2}$~s$^{-1}$) which contribute most
significantly to the XRB. 

The wider \hbox{X-ray} surveys provide a broad census of the \hbox{X-ray} source
population, often generating enormous numbers of sources (\hbox{1000--6000} or 
more; e.g., see Figure~2). They thereby allow discovery 
of both intrinsically rare source types
as well as low-redshift examples of sources found in the deep \hbox{X-ray} surveys. 
However, complete multiwavelength follow-up often must be compromised for 
reasons of observational economy; thus many of the wider surveys target 
specific source types of interest. Often targeted are sources with unusually
hard \hbox{X-ray} spectra, sources with unusually large X-ray-to-optical flux ratios, 
or sources that appear likely to be at high redshift based upon optical
imaging data. 

\begin{figure}[t!]
\centering
\includegraphics[height=10cm]{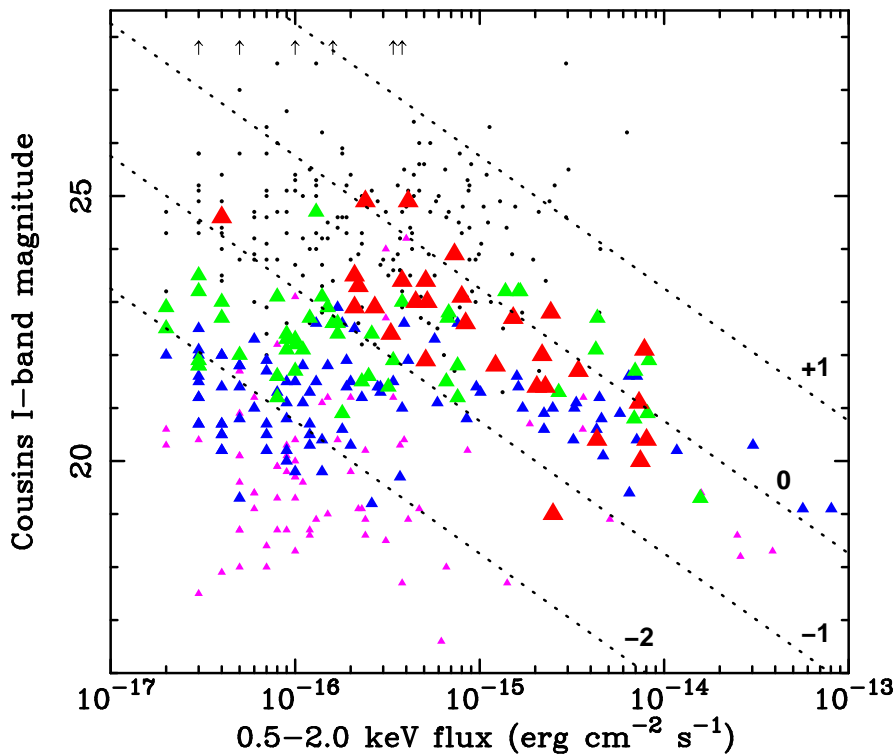}
\caption{$I$-band magnitude versus \hbox{0.5--2~keV} flux for extragalactic \hbox{X-ray}
sources in the \hbox{CDF-N}. Sources with redshifts of \hbox{0--0.5}, \hbox{0.5--1}, 
\hbox{1--2}, and \hbox{2--6} are shown as violet, blue, green, and red filled 
triangles, respectively (symbol sizes also increase with redshift). 
Small black dots indicate sources without measured redshifts. The slanted, dotted
lines indicate constant values of $\log (f_{\rm X}/f_{\rm I})$; the respective
$\log (f_{\rm X}/f_{\rm I})$ values are labeled. Adapted from 
D.M. Alexander, F.E. Bauer, W.N. Brandt, et~al., 2003, AJ, 126, 539 and 
A.J. Barger, L.L. Cowie, P. Capak, et~al., 2003, AJ, 126, 632.}
\label{fig:3}       
\end{figure}

% --------------------------------------------------------------------------------

\section{Some Implications for AGN Physics and Evolution}
\label{sec:3}

\subsection{Properties of the X-ray Sources}
\label{sec:3p1}

\subsubsection{Basic Nature}
\label{sec:3p1p1}

A broad diversity of \hbox{X-ray} sources is found in the recent \chandra\ and 
\xmm\ surveys. This is apparent in even basic flux-flux plots such as that
shown in Figure~3; at the faintest \hbox{X-ray} flux levels in the \hbox{CDF-N}, the 
extragalactic sources range in optical flux by a factor of $\simgt 10,000$. 

Classification of the \hbox{X-ray} sources is challenging for several reasons. 
First of all, many of the sources are simply too faint for efficient optical 
spectroscopic identification with 8--10~m class 
telescopes (note the small dots in Figure~3).
Intensive optical identification programs on the deepest \chandra\ and \xmm\
fields typically have \hbox{$\approx 50$--70\%} completeness at best. 
Furthermore, many of the \hbox{X-ray} sources have modest apparent
optical luminosities, and thus their host galaxies make substantial 
diluting contributions to the flux measured in a spectroscopic aperture. 
Finally, another challenge is an apparent ``schism'' between optical 
(type~1 vs. type~2) and \hbox{X-ray} (unobscured vs. obscured) schemes of
classification; not all \hbox{X-ray} obscured AGN have type~2 optical spectra, and 
not all AGN with type~1 optical spectra are unobscured. 

\begin{figure}[t!]
\centering
\includegraphics[height=10cm]{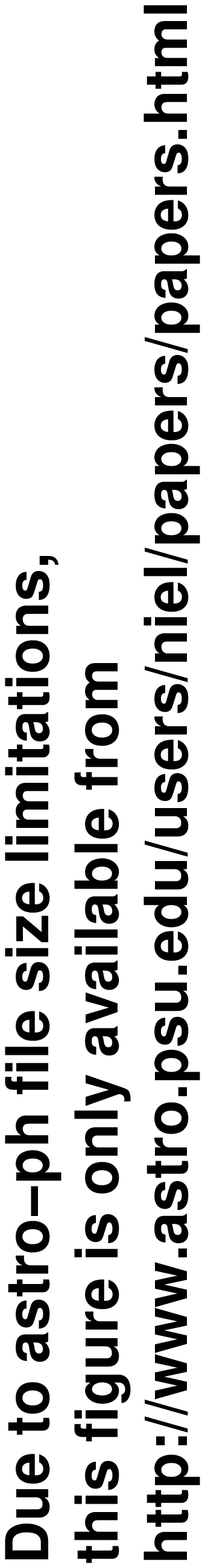}
\caption{\chandra\ and \hst\ images of the HDF-N. The 22 \chandra\ sources 
are circled on the \hst\ image; the circles are much larger than the \chandra\ 
source positional errors. The numbers are source redshifts; redshifts followed
by a ``p'' are photometric. Basic source type information for many of the
sources is also given.}
\label{fig:4}       
\end{figure}

Considering \hbox{X-ray}, optical, and multiwavelength information, the
primary extragalactic source types are found to be the following: 

\begin{description}

\item{$\bullet$}
{\it Unobscured AGN.\/} 
Blue, broad-line AGN are found that do not show
signs of obscuration at either \hbox{X-ray} or optical/UV wavelengths. They
are found over a broad range of redshift \hbox{($z\approx 0$--5)}, and they
comprise a significant fraction of the brightest X-ray sources. At 
$z\simgt 1.5$ they also comprise a substantial fraction of all \hbox{X-ray} 
sources with spectroscopic identifications (certainly in part because these
objects are the most straightforward to identify spectroscopically). 

\item{$\bullet$}
{\it Obscured AGN with clear optical/UV AGN signatures.\/} 
Some objects showing \hbox{X-ray} evidence for obscuration have clear AGN 
signatures in their rest-frame optical/UV spectra. Notably, such AGN 
can have both type~1 and type~2 optical/UV classifications. Most of 
these objects have $z\simlt 1.5$. 

\item{$\bullet$}
{\it Optically faint X-ray sources.\/} 
These sources have $I\simgt 24$ and thus usually cannot be identified 
spectroscopically. Many, however, appear to be luminous, obscured 
AGN at \hbox{$z\approx 1$--3} when their \hbox{X-ray} properties, optical 
photometric properties (including photometric redshifts), and 
multiwavelength properties are considered. Thus, these objects largely 
represent an extension of the previous class to higher redshifts and 
fainter optical magnitudes. 

\item{$\bullet$}
{\it X-ray bright, optically normal galaxies (XBONGs).\/} 
XBONGs have \hbox{X-ray} luminosities ($\approx 10^{41}$--$10^{43}$~erg~s$^{-1}$)
and X-ray-to-optical flux ratios suggesting some type of 
moderate-strength AGN activity. Some also have hard \hbox{X-ray} spectral
shapes suggesting the presence of \hbox{X-ray} obscuration. Optical spectra
give redshifts of \hbox{$z\approx 0.05$--1}, but AGN emission lines
and non-stellar continua are not apparent. 
The nature of XBONGs remains somewhat mysterious. Some may 
just be Seyfert~2s where dilution by host-galaxy light 
hinders optical detection of the AGN, but others have high-quality 
follow up and appear to be truly remarkable objects. These 
``true'' XBONGs may be
(1) AGN with inner radiatively inefficient accretion flows, or 
(2) AGN that suffer from heavy obscuration covering a large solid 
angle ($\approx 4\pi$~sr), so that optical emission-line 
and ionizing photons cannot escape the nuclear region. 

\item{$\bullet$}
{\it Starburst galaxies.\/}
At the faintest \hbox{X-ray} flux levels in the deepest \chandra\ surveys, a 
significant fraction of the detected sources appear to be \hbox{$z\approx 0$--1.3} 
dusty starburst galaxies. They are members of the strongly evolving starburst 
population responsible for creating much of the infrared background. 
The observed \hbox{X-ray} flux appears to be the integrated emission from many
\hbox{X-ray} binaries and supernova remnants. 

\item{$\bullet$}
{\it ``Normal'' galaxies.\/}
Apparently normal galaxies are also detected in the deepest \chandra\ surveys 
out to $z\approx 0.2$. The observed \hbox{X-ray} emission is again probably largely 
from \hbox{X-ray} binaries and supernova remnants; these objects and the
starburst galaxies above are probably not distinct but rather constitute
a single population of galaxies with star formation of varying intensity. 
Low-luminosity AGN are likely present in some cases as well. Some normal galaxies 
sport luminous \hbox{X-ray} sources clearly offset from their nuclei. At even 
fainter \hbox{X-ray} flux levels, normal and starburst galaxies should be the 
dominant class of extragalactic \hbox{X-ray} sources. 

\end{description}

\noindent
Most of the AGN found in X-ray surveys are ``radio quiet'' in the
sense that the ratio $(R)$ of their rest-frame 5~GHz and 4400~\AA\ flux
densities are $R<10$. 

Figure~4 shows some of the source classifications in the HDF-N, which is
at the center of the \hbox{CDF-N} (see Figure~1) and thus has the most sensitive
\hbox{X-ray} coverage available. Note, for example, that three of the brightest 
\hbox{X-ray} sources are XBONGs. These were not recognized as AGN prior to the 
\chandra\ observations, despite the many intensive studies of the HDF-N. 

\subsubsection{Luminosity and Redshift Distributions}
\label{sec:3p1p2}

The combined results from deep and wider \hbox{X-ray} surveys show that the sources
comprising most of the XRB have \hbox{X-ray} luminosities comparable to those of
local Seyfert galaxies, such as NGC~3783, NGC~4051, 
and NGC~5548 (e.g., see Figure~5). While a few remarkable 
obscured quasars have been found, these appear fairly rare and
only make a small contribution to the XRB. Indeed, it appears that the
fraction of obscured AGN drops with luminosity from \hbox{$\approx 60$--70\%} 
at Seyfert luminosities to $\approx 30$\% at quasar luminosities. 

\begin{figure}[t!]
\centering
\includegraphics[height=9cm]{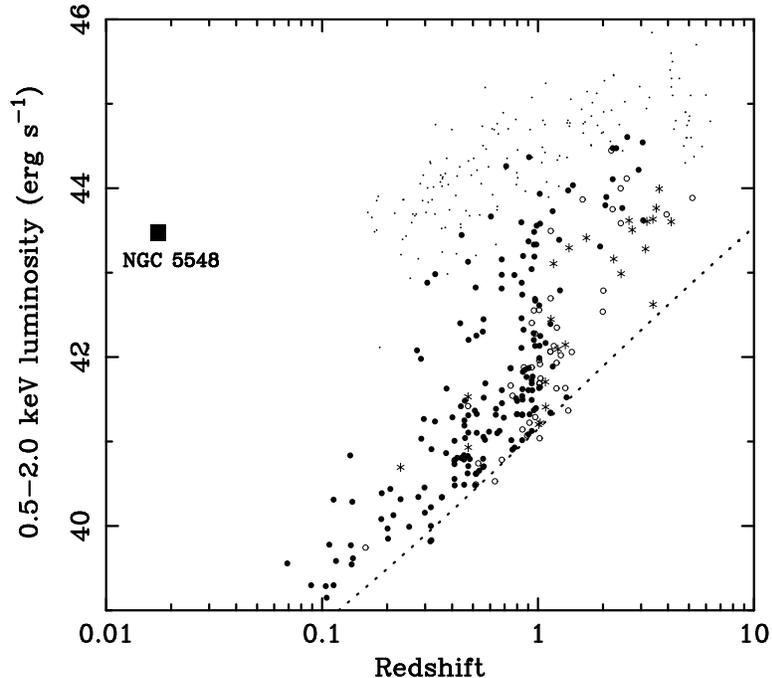}
\caption{Luminosity in the \hbox{0.5--2~keV} band (computed from the 0.5--2~keV
flux assuming a power-law spectrum with a photon index of $\Gamma=2$) 
versus redshift for extragalactic sources in 
the \hbox{CDF-N} with spectroscopic redshifts. Sources with
\hbox{$I=16$--22}, \hbox{$I=22$--23}, and $I>23$ are indicated with filled circles, 
open circles, and stars, respectively. The dotted curve shows the approximate 
sensitivity limit near the center of the \hbox{CDF-N}. Also shown are 
the well-studied Seyfert~1 galaxy NGC~5548 (filled square) and 
Sloan Digital Sky Survey (SDSS) quasars from the SDSS Early Data
Release with \hbox{X-ray} coverage in archival \rosat\ data (small dots; the relevant 
solid angle covered by pointed \rosat\ observations is $\approx 15$~deg$^{2}$). 
Note that NGC~5548 could have been detected to $z\sim 10$ in the \hbox{CDF-N}. 
Note also that the \hbox{CDF-N} and SDSS populations are nearly disjoint, as a
consequence of the different solid angle coverages (a factor of
$\sim 120$) and depths.
Adapted from D.M. Alexander, F.E. Bauer, W.N. Brandt, et~al., 2003, AJ, 126, 539; 
A.J. Barger, L.L. Cowie, P. Capak, et~al., 2003, AJ, 126, 632; and
C. Vignali, W.N. Brandt, \& D.P. Schneider, 2003, AJ, 125, 433.}
\label{fig:5}       
\end{figure}

Most spectroscopically identified AGN in the deep \hbox{X-ray} surveys have 
$z\simlt 2$, although a significant minority have \hbox{$z\approx 2$--5}. 
This is partly due to spectroscopic incompleteness bias, as is apparent
by comparing the filled circles, open circles, and stars in Figure~5. 
However, as will be described further in \S3.2, there is a real enhancement
of AGN at $z\simlt 1$ relative to expectations from pre-\chandra\ AGN-synthesis
models of the XRB. An impressive $\sim 60$\%\ of the 2--8~keV XRB arises at 
$z<1$. 

\subsubsection{AGN Sky Density}
\label{sec:3p1p3}

Most \hbox{($\approx 70$--100\%)} of the 
extragalactic \hbox{X-ray} sources found in both the 
deep and wider \hbox{X-ray} surveys 
with \chandra\ and \xmm\ are AGN of some type. Starburst and 
normal galaxies make increasing fractional contributions at the faintest \hbox{X-ray} 
flux levels, but even in the \hbox{CDF-N} they represent \hbox{$\simlt 20$--30\%} 
of all sources (and create $\simlt 5$\%\ of the XRB). 
The observed AGN sky density in the deepest \hbox{X-ray} surveys is 
$\approx 6500$~deg$^{-2}$, about an order of magnitude higher than that
found at any other wavelength. This exceptional effectiveness at finding
AGN arises because \hbox{X-ray} selection (1) has reduced absorption bias and 
minimal dilution by host-galaxy starlight, and (2) allows concentration 
of intensive optical spectroscopic follow-up upon high-probability 
AGN with faint optical counterparts (i.e., it is possible to 
probe further down the luminosity function). 

\subsubsection{Completeness of X-ray AGN Selection}
\label{sec:3p1p4}

Are there significant numbers of luminous AGN that are not found even
in the deepest \hbox{X-ray} surveys? This could be the case if there is a 
large population of AGN that are \hbox{X-ray} weak due either to absorption
or an intrinsic inability to produce \hbox{X-rays}. This question can be
partially addressed by looking for AGN found at other wavelengths that
are not detected in \hbox{X-rays}. In the \hbox{CDF-N}, one of the most intensively
studied regions of sky at all wavelengths, there are only \hbox{1--2} such AGN
known. The most conspicuous is 123725.7+621128, a radio-bright 
($\approx 6$~mJy at 1.4~GHz) wide angle tail source that is likely
at \hbox{$z\approx 1$--2} (although the redshift of this source remains uncertain). 
This is one of the brightest radio sources in the \hbox{CDF-N} but has been 
notoriously difficult to detect in \hbox{X-rays}. Manual analysis of the 
2~Ms \chandra\ data at the AGN position indicates a likely, but still 
not totally secure, detection (at a false-positive probability threshold 
of $3\times 10^{-5}$ using the standard \chandra\ wavelet source 
detection algorithm). The \hbox{0.5--2~keV} luminosity is 
$\simlt 5\times 10^{41}$~erg~s$^{-1}$. The only other known AGN in 
the \hbox{CDF-N} without an \hbox{X-ray} detection is 123720.0+621222, 
a narrow-line AGN at $z=2.45$; its \hbox{0.5--2~keV} luminosity is 
$\simlt 2\times 10^{42}$~erg~s$^{-1}$. 

Despite the spectacular success of \hbox{X-ray} surveys at finding AGN, 
appropriate humility is required when assessing the AGN selection
completeness of even the deepest \hbox{X-ray} surveys. This is made clear by 
consideration of ``Compton-thick'' AGN, which comprise a sizable 
fraction ($\approx 40$\%) of AGN in the local universe. Such AGN are 
absorbed by intrinsic column densities of 
$N_{\rm H}\gg 1.5\times 10^{24}$~cm$^{-2}$, within which direct 
line-of-sight \hbox{X-rays} are effectively destroyed via the combination 
of Compton scattering and photoelectric absorption. Such AGN are often 
only visible via weaker, indirect \hbox{X-rays} that are ``reflected'' by 
neutral material or ``scattered'' by ionized material.\footnote{In some
``translucent'' cases, where the column density is only a few 
$\times 10^{24}$~cm$^{-2}$ (i.e., a few Thomson depths), direct 
``transmission'' X-rays from a Compton-thick AGN may become visible above 
rest-frame energies of $\sim 10$~keV. For comparison, the column density 
through your chest is $\sim 1\times 10^{24}$~cm$^{-2}$; if you stood
along the line-of-sight to an AGN, you could almost render it Compton thick!} Many 
of the local Compton-thick AGN (e.g., NGC~1068, NGC~6240, Mrk~231), if placed
at \hbox{$z\simgt 0.5$--1.5}, would remain undetected in even the deepest \chandra\
surveys. Thus, it appears plausible that $\approx 40$\% of AGN at such 
redshifts may have been missed (the number, of course, could be higher
or lower if the fraction of Compton-thick AGN evolves significantly
with redshift). Deeper observations with \chandra\ 
($\approx 10$~Ms; see \S4.1) may be able to detect the indirect \hbox{X-rays} from 
any missed Compton-thick AGN, and observations with \spitzer\ may be able 
to detect ``waste heat'' from such objects at infrared wavelengths. 

Another way to address AGN selection completeness in \hbox{X-ray} surveys is
to consider ``book-keeping'' arguments: can the observed sources
explain the observed \hbox{20--40~keV} XRB intensity, and can all
the observed accretion account for the local density of
supermassive black holes? The answer is plausibly ``yes'' in both cases, 
but with some uncertainty. In the first case, one must make a
significant spectral extrapolation from \hbox{5--10~keV} and worry 
about mission-to-mission cross-calibration uncertainties. In the second, 
significant uncertainties remain in bolometric correction factors,
accretion efficiencies, and the local density of supermassive black
holes. The current book-keeping arguments cannot rule out the possibility 
that a significant fraction of the AGN population (e.g., Compton-thick AGN)
is still missed in X-ray surveys. Indeed, some book-keepers find better
agreement with the local black-hole mass function after making a substantial 
correction for missed accretion in Compton-thick AGN. 

% --------------------------------------------------------------------------------

\subsection{Recent X-ray Results on AGN Evolution}
\label{sec:3p2}

\begin{figure}[t!]
\centering
\includegraphics[height=5.8cm,width=5.8cm]{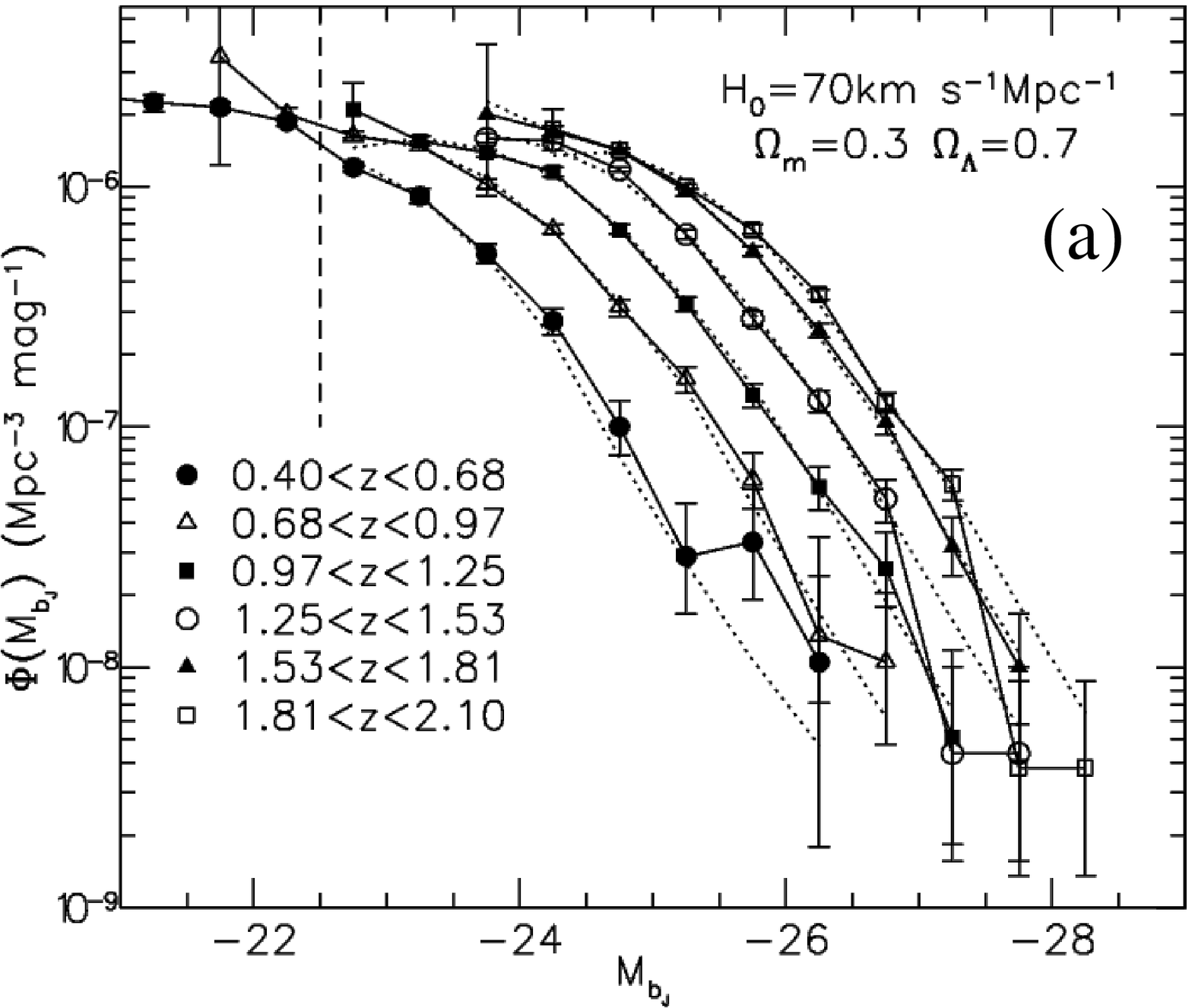}
\includegraphics[height=5.8cm]{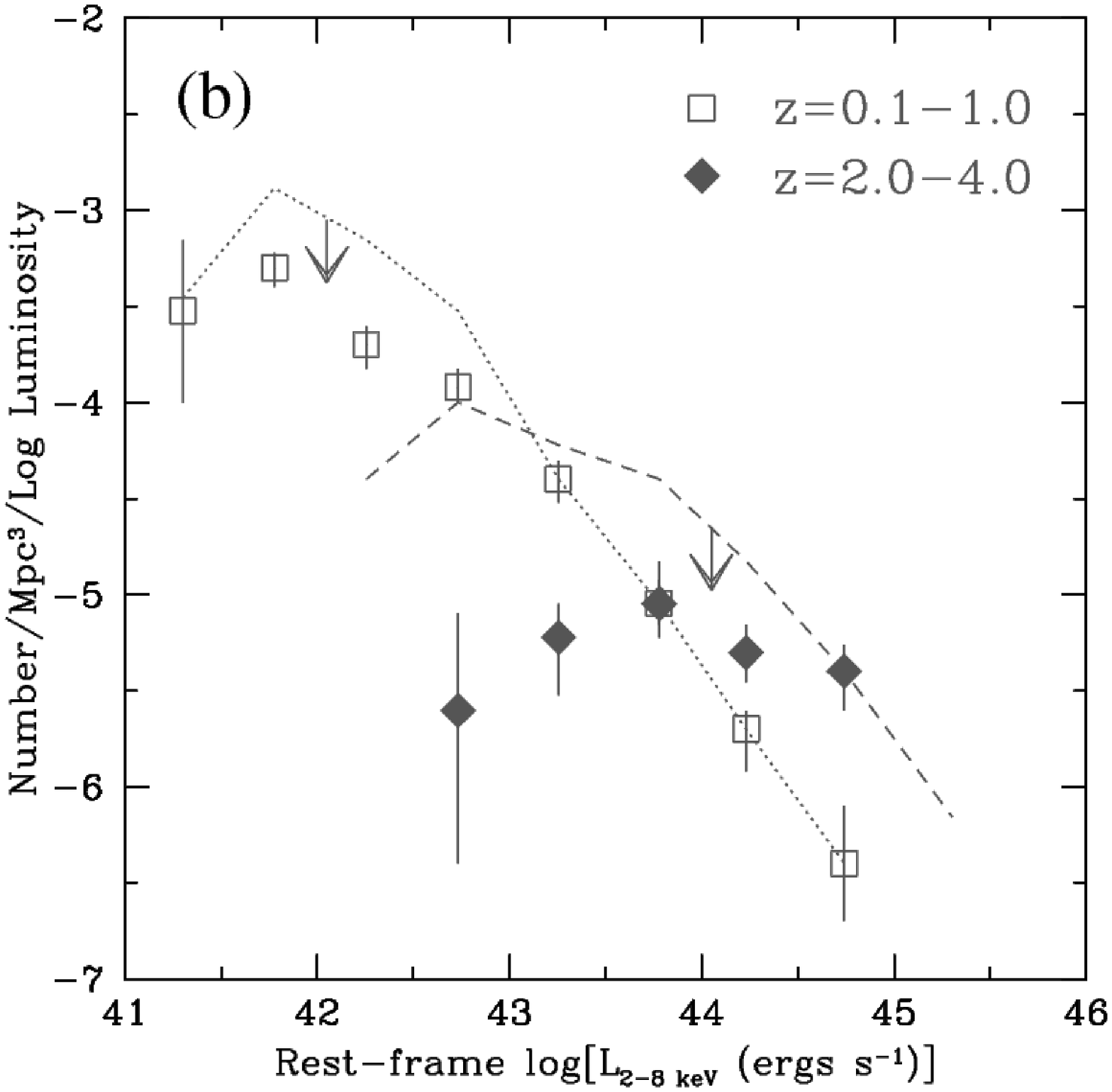}
\caption{(a) Optical luminosity functions in 6 redshift ``shells'' spanning
\hbox{$z=0.40$--2.10} for $\sim 16,800$ luminous AGN from the 2dF and 6dF 
surveys. Note the clear positive evolution with increasing redshift at 
high luminosity (i.e., the comoving number density of luminous AGN 
increases with redshift from \hbox{$z=0.40$--2.10}). 
From S.M. Croom, R.J. Smith, B.J. Boyle, et~al., 2004, MNRAS, 
in press (astro-ph/0403040). 
(b) \hbox{X-ray} \hbox{(2--8~keV)} luminosity functions in two redshift ``shells'' (as
labeled) for moderate-to-high luminosity AGN from the \hbox{CDF-N}, Abell 370, 
SSA13, and SSA22 \chandra\ surveys (see Table~1) as well as several 
earlier \hbox{X-ray} surveys. The dotted and dashed curves show the maximum
possible luminosity functions after allowing for incompleteness
of the follow-up spectroscopy. Note the apparent negative evolution with 
increasing redshift at moderate luminosity. Adapted from 
L.L. Cowie, G.P. Garmire, M.W. Bautz, et~al., 2002, ApJ, 566, L5.}
\label{fig:6}       
\end{figure}

Optical studies of AGN evolution have typically focused on luminous quasars. 
These have been known to evolve strongly with redshift since $\sim 1968$,
having a comoving space density at $z\approx 2$ that is $\simgt 100$ 
times higher than at $z\approx 0$. Figure~6a shows optical luminosity
functions in 6 redshift ``shells'' spanning \hbox{$z=0.40$--2.10} for 
$\sim 16,800$ luminous AGN from the 2dF and 6dF surveys. Clear positive 
evolution with redshift is observed, and pure luminosity evolution (PLE) 
models provide an acceptable fit to these data. 
New optical AGN surveys, such as COMBO-17, have recently discovered 
significant numbers of moderate-luminosity AGN (with $M_{\rm B}>-23$) 
at \hbox{$z\approx 1$--4}, allowing investigation of their evolution. As for 
luminous quasars, the AGN found in these surveys also appear to peak in 
comoving space density at $z\approx 2$. Both PLE and pure 
density evolution (PDE) models can acceptably fit the COMBO-17 data 
alone. Although a systematic combination of the COMBO-17 data with a 
large sample of higher-luminosity AGN has yet to be published, there are
hints that the redshift at which the comoving space density peaks is 
smaller at lower luminosities. 

\begin{figure}[t!]
\centering
\includegraphics[height=9cm]{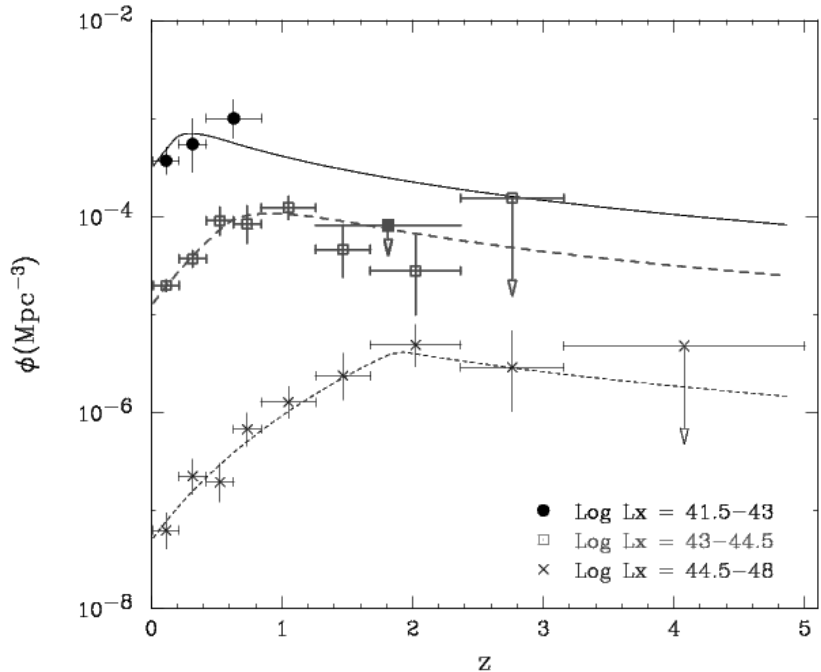}
\caption{The comoving spatial density of AGN in three \hbox{X-ray} luminosity
ranges as a function of redshift, derived using data from several \hbox{X-ray}
surveys. From Y. Ueda, M. Akiyama, K. Ohta, et~al., 2003, ApJ, 598, 886.}
\label{fig:7}       
\end{figure}

As noted in \S3.1, the deepest \hbox{X-ray} surveys efficiently select AGN
even fainter than those found by COMBO-17 out to high redshift
(e.g., see Figure~5). \hbox{X-ray} AGN samples show a clear dependence
of AGN evolution upon luminosity, with strong positive evolution only 
being seen at high luminosities (see Figure~6b). Lower luminosity 
AGN appear to be about as common at \hbox{$z\approx 0$--1} as they ever
were, consistent with trend hinted at by COMBO-17. These results are robust 
to incompleteness of the spectroscopic follow up, although clearly they 
are still dependent upon the completeness of AGN \hbox{X-ray} selection 
(see \S3.1). It appears that while the SMBH in rare, luminous AGN could 
grow efficiently at high redshift, the SMBH in most AGN had to wait 
longer to grow. 

Figure~7 shows estimates of the comoving spatial density of AGN in 
three \hbox{X-ray} luminosity ranges as a function of redshift. These have
been constructed utilizing a combination of \chandra, \asca, and 
\heao1\ surveys at photon energies above 2~keV (with 247 AGN in total). 
The data are best fit with luminosity-dependent density evolution 
(LDDE) out to some cutoff redshift ($z_{\rm c}$), where $z_{\rm c}$ increases
with luminosity; as a result, the ratio of the peak spatial density to
that at the present day is higher for more luminous AGN. At a basic
level, LDDE also seems more physically plausible than PLE or PDE; simple
PLE models tend to overpredict the number of $\simgt 10^{10}$~M$_\odot$ 
black holes in the local universe, while simple PDE models tend to 
overpredict the local space density of quasars.

% --------------------------------------------------------------------------------

\subsection{X-ray Emitting AGN in Luminous Submillimeter Galaxies}
\label{sec:3p3}

\begin{figure}[t!]
\centering
\includegraphics[height=9cm]{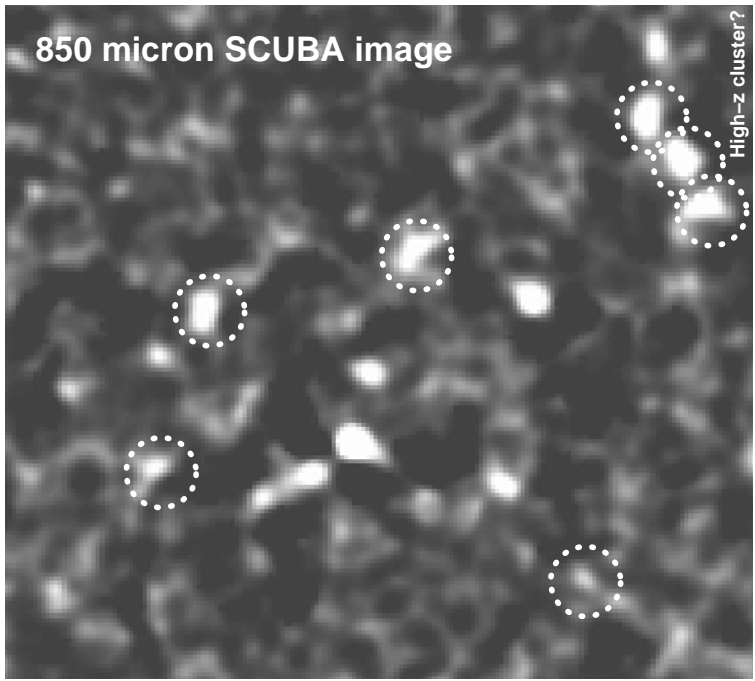}
\caption{Map at 850~$\mu$m of the central region of the \hbox{CDF-N}; the map is 
$\approx 6^\prime$ on a side. Sources at 850~$\mu$m with \hbox{X-ray} detections 
are enclosed by dotted circles. The three clustered 850~$\mu$m/X-ray 
sources near the upper-right corner are also coincident with an extended \hbox{X-ray}
source, perhaps a high-redshift cluster. Adapted from 
D.M. Alexander, F.E. Bauer, W.N. Brandt, et~al., 2003, AJ, 125, 383 and 
C. Borys, S. Chapman, M. Halpern, et~al., 2003, MNRAS, 344, 385.}
\label{fig:8}       
\end{figure}

The deepest \chandra\ and \xmm\ surveys have finally provided the 
necessary \hbox{X-ray} sensitivity to complement the most sensitive surveys 
at submillimeter and infrared wavelengths. One notable instance 
where obtaining the highest possible \hbox{X-ray} sensitivity has been 
essential is in studies of the AGN content of distant 
submillimeter galaxies detected with the SCUBA instrument on the
James Clerk Maxwell Telescope. Most of these galaxies
are thought to contain intense starbursts with star-formation 
rates of \hbox{$\simgt 100$~M$_\odot$~yr$^{-1}$}, yet they are 
not notable in optical galaxy surveys due to dust obscuration 
of the corresponding starlight. The SCUBA galaxy population 
is thought to be mostly at \hbox{$z\approx 1.5$--3}, and such galaxies 
were $\sim 1000$ times more common at $z\sim 2$ than in the local
universe. The obscured starlight in submillimeter galaxies is
re-radiated in the rest-frame infrared (observed-frame submillimeter).

% , and integrated over the lifetime of the Universe 
% galaxies have radiated comparable amounts of energy at 
% infrared/submillimeter and optical wavelengths. 

\begin{figure}[t!]
\centering
\includegraphics[height=10cm]{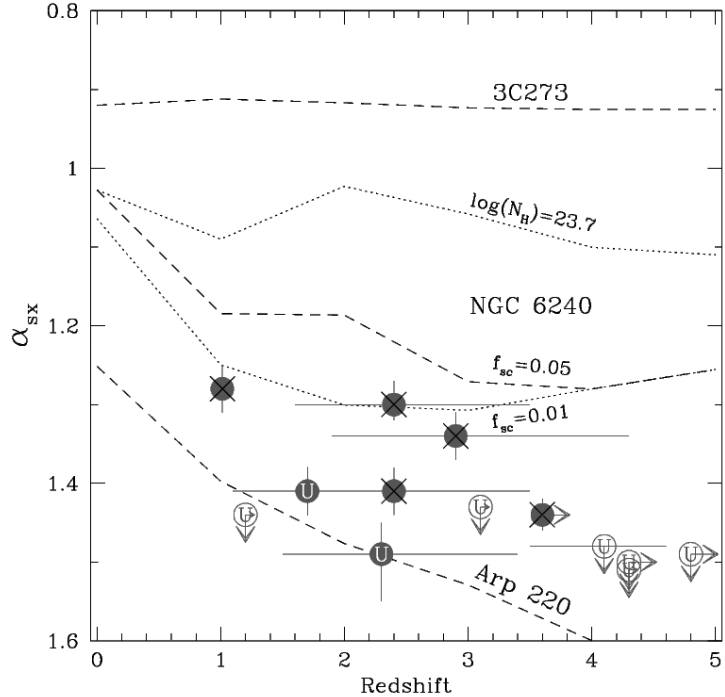}
\caption{Submillimeter-to-X-ray spectral index ($\alpha_{\rm sx}$) versus
redshift. Submillimeter sources in the central part of the \hbox{CDF-N} 
with (without) \hbox{X-ray} detections are shown as solid (open) 
circles. The five circles with overlaid crosses are likely AGN 
according to their X-ray properties, while those with overlaid ``U'' are 
of unknown X-ray type. Dashed curves show
$\alpha_{\rm sx}$ values for 3C~273, NGC~6240, and 
Arp~220 adopting their observed amounts of X-ray absorption; alternative
dotted curves show an AGN like NGC~6240 but with less internal absorption 
($N_{\rm H}=5\times 10^{23}$~cm$^{-2}$) and a smaller scattered
flux fraction ($f_{\rm sc}=0.01$). Adapted from 
D.M. Alexander, F.E. Bauer, W.N. Brandt, et~al., 2003, AJ, 125, 383.}
\label{fig:9}       
\end{figure}

What fraction of submillimeter galaxies contains actively 
accreting supermassive black holes? Sensitive \hbox{X-ray} 
studies play an important role in addressing this
question, since they allow effective searching for AGN in the 
majority of submillimeter galaxies that are optically faint (and 
thus challenging to study in detail with optical spectroscopy). 
Early comparisons between \hbox{$\approx 20$--150~ks} \chandra\ surveys 
and submillimeter surveys yielded little ($\simlt 10$\%) source 
overlap. However, the latest analysis of the 2~Ms \hbox{CDF-N} data
reveals that seven of the 13 ($\approx 54$\%) bright submillimeter 
galaxies (with 850~$\mu$m flux densities of $>5$~mJy)
in the \hbox{CDF-N} central region mapped with SCUBA have \hbox{X-ray} 
counterparts (see Figure~8); these counterparts have 
\hbox{$\approx 15$--200} counts in the full \chandra\ bandpass. 
Five of the seven \hbox{X-ray} detected submillimeter galaxies likely
host obscured AGN based upon their observed \hbox{X-ray} luminosities, 
\hbox{X-ray} spectral shapes, and X-ray-to-submillimeter flux ratios
(see Figure~9). The remaining two have \hbox{X-ray} emission properties
consistent with those expected from star formation activity, 
although it is possible that they host weak AGN as well. 
If the latter two sources are indeed powered mainly by star 
formation, they would be the most \hbox{X-ray} luminous 
($\approx 4\times 10^{42}$~erg~s$^{-1}$) starburst galaxies 
known. 

Do the \hbox{X-ray} emitting AGN found in many submillimeter galaxies make 
a significant contribution to these galaxies' total energy output? 
Answering this question requires assessment of the amount of \hbox{X-ray} 
absorption present since, for a given observed \hbox{X-ray} flux, a 
Compton-thick AGN can be much more luminous than a Compton-thin AGN
(see \S3.1). Basic \hbox{X-ray} spectral fitting suggests that three of the 
five submillimeter galaxies hosting AGN in the \hbox{CDF-N} central region 
have Compton-thin absorption, while only one is likely to have 
Compton-thick absorption (the final object has poor \hbox{X-ray} spectral
constraints). Armed with this knowledge, consideration of the
observed \hbox{X-ray-to-submillimeter} flux ratios (see Figure~9) suggests
that $\simlt 10$\% of the total energy output from these submillimeter 
galaxies is ultimately due to an AGN. Star-formation is apparently 
the dominant power source for the infrared/submillimeter emission, 
even when an AGN is also present. 

The results above are currently being extended, utilizing redshifts
from ongoing deep optical spectroscopy. Thus far, these extended
results confirm the main conclusions above. 

% --------------------------------------------------------------------------------

\subsection{High-Redshift (\boldmath$z>4$) AGN Demography and Physics}
\label{sec:3p4}

As is apparent from Figures~5 and 10, 
deep \hbox{X-ray} surveys can detect $z>4$ AGN that
are $\simgt 10$--30 times less luminous than those found in wide-field optical 
AGN surveys such as the SDSS. At least in the local universe, such 
moderate-luminosity AGN are much more numerous and thus more 
representative than the rare, highly luminous quasars. Furthermore, unlike 
the rest-frame ultraviolet light sampled at $z>4$ in ground-based AGN 
surveys, \hbox{X-ray} surveys suffer from progressively {\it less\/} absorption  
bias as higher redshifts are surveyed. At $z>4$, hard $\approx 2$--40~keV 
rest-frame \hbox{X-rays} are accessed; these can penetrate large column
densities up to several $\times 10^{24}$~cm$^{-2}$. 

\begin{figure}[t!]
\centering
\includegraphics[height=10cm]{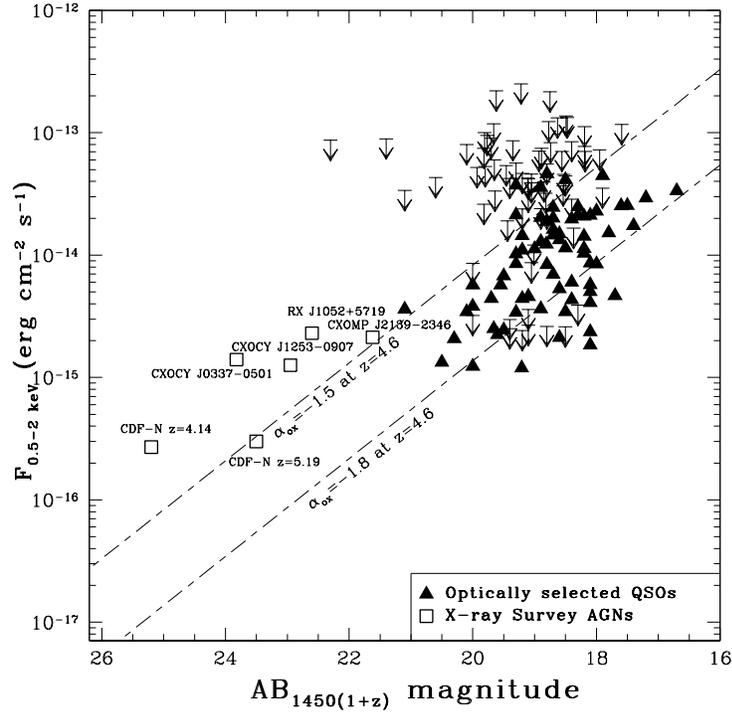}
\caption{Observed-frame, Galactic absorption-corrected \hbox{0.5--2~keV} flux 
versus AB$_{1450(1+z)}$ magnitude for $z\ge4$ AGN found both in optical and
\hbox{X-ray} surveys (the X-ray upper limits shown are all for AGN from optical
surveys). The slanted, dashed lines show the 
$\alpha_{\rm ox}=-1.5$ and $\alpha_{\rm ox}=-1.8$ loci at $z=4.6$. 
Adapted from C. Vignali, W.N. Brandt, D.P. Schneider, et~al., 2003, AJ, 125, 2876.}
\label{fig:10}       
\end{figure}

Spectroscopic follow-up of moderate-luminosity \hbox{X-ray} detected AGN at $z>4$ 
is challenging, since such objects are expected to have $z$ magnitudes of 
23--26 (provided they have not ``dropped out'' of the $z$ bandpass entirely). 
Nevertheless, significant constraints on the sky density of such objects
have been set via large-telescope spectroscopy and Lyman-break selection. 
In the latter case, objects can be selected that either have appropriate
optical/near-infrared colors to be at $z>4$ or alternatively have
no optical/near-infrared detections. The ``bottom line'' from these 
demographic studies in the \hbox{CDF-N} and \hbox{CDF-S} is that there are $\simlt 12$ 
AGN at $z>4$ detectable in a 1--2~Ms \chandra\ field, and 
that only $\approx 4$ of these have a $z$ magnitude of $<25$ (this 
limit on the sky density is still $\sim 260$ times the sky density
of $z>4$ quasars from the SDSS). These sky-density constraints 
are sufficient to rule out some pre-\chandra\ predictions by 
about an order of magnitude, and the combined \hbox{X-ray} and
SDSS results indicate that the AGN contribution to reionization at 
$z\approx 6$ is small. 

\begin{figure}[t!]
\centering
\includegraphics[height=5.0cm]{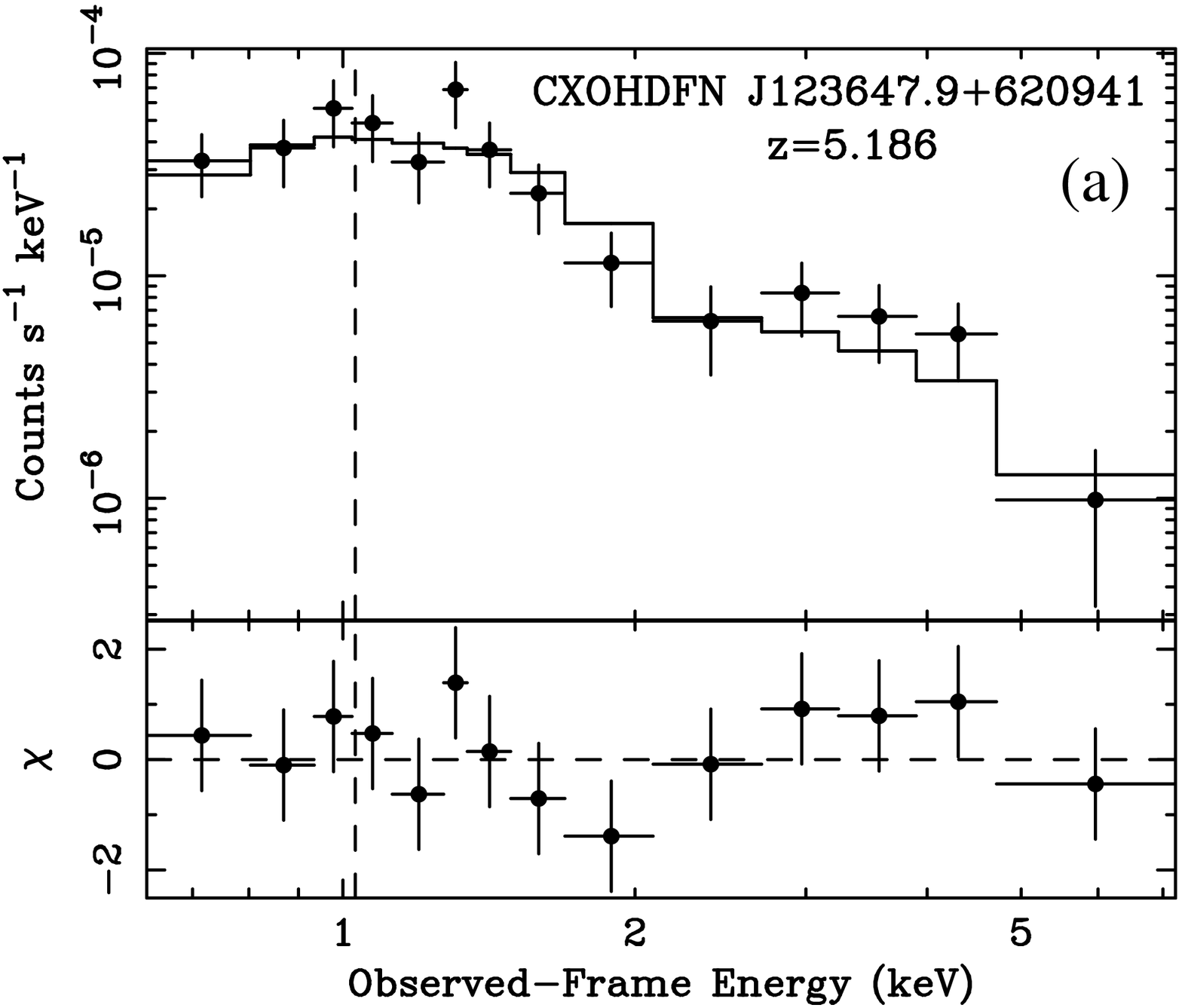}
\includegraphics[height=5.0cm]{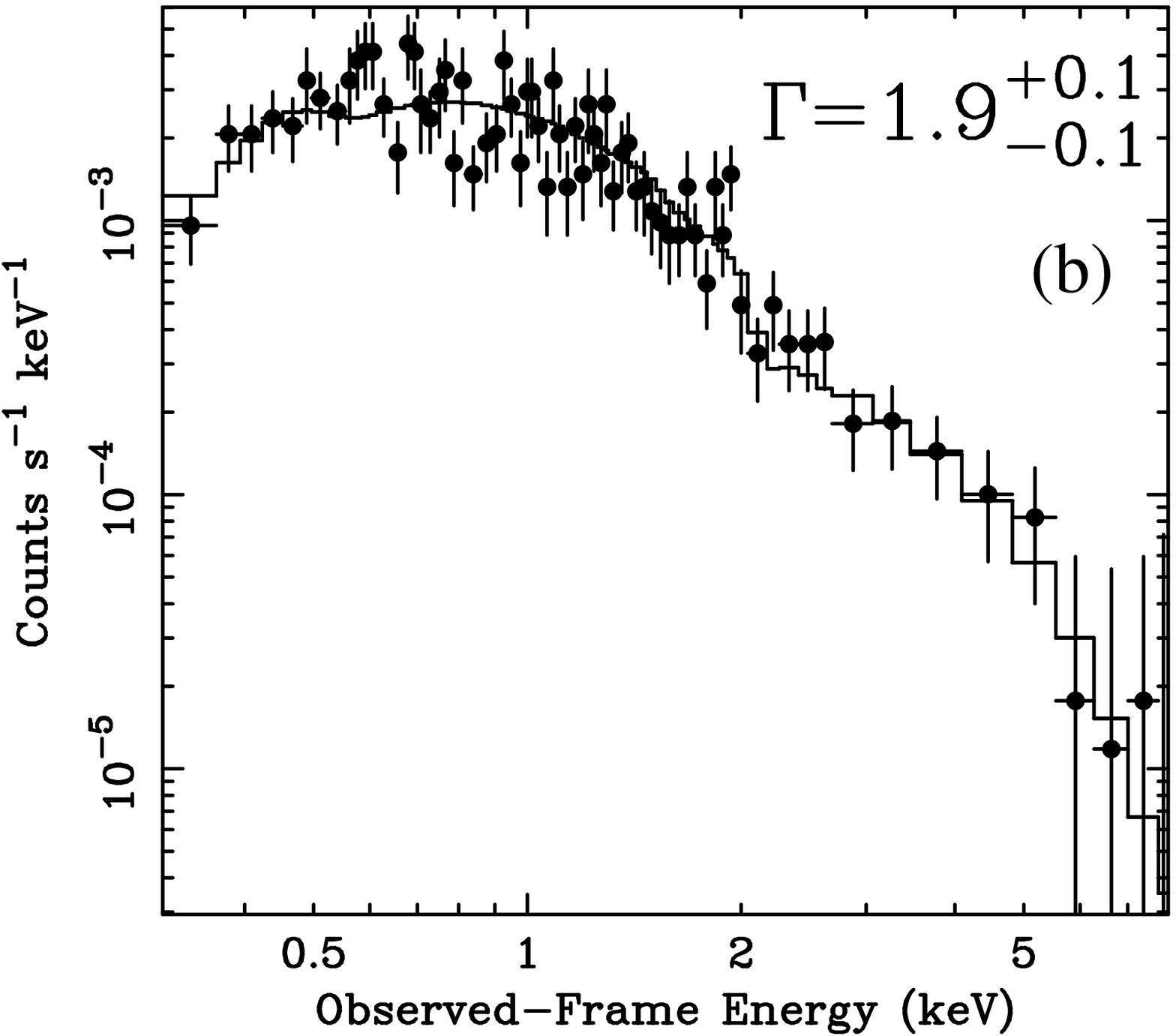}
\caption{Observed-frame \hbox{X-ray} spectra for
(a) the $z=5.186$ \hbox{CDF-N} AGN CXOHDFN~J123647.9+620941 and 
(b) 46 radio-quiet quasars at \hbox{$z=4.0$--6.3} that have been stacked together. 
The best-fitting power-law models with Galactic absorption are also shown; 
see the text for fitting results. In (a) the lower panel shows the fit
residuals in units of sigma, and the vertical dashed line indicates
the energy of the (undetected) 6.4~keV iron~K$\alpha$ line. Adapted from 
C. Vignali, F.E. Bauer, D.M. Alexander, et~al., 2002, ApJ, 580, L105 and 
C. Vignali, W.N. Brandt, \& D.P. Schneider, 2004, astro-ph/0310659.}
\label{fig:11}       
\end{figure}

Once high-redshift AGN have been identified, via either \hbox{X-ray} or 
optical surveys, broad-band spectral energy distribution 
analyses and \hbox{X-ray} spectral fitting can provide 
information on their accretion processes and 
environments. The currently available data, albeit limited, suggest
that $z>4$ AGN are accreting and growing in roughly the same
way as AGN in the local universe; there is no evidence that their
inner \hbox{X-ray} emitting regions have been affected by, for example, 
accretion-disk instabilities or radiation-trapping effects. 
Figure~10 plots \hbox{X-ray} versus optical flux for $z>4$ AGN from 
both \hbox{X-ray} and optical surveys. The X-ray-to-optical spectral
indices, $\alpha_{\rm ox}$, for these objects are consistent with those 
of AGN in the local universe, once luminosity effects and selection 
biases are taken into account. These biases and effects likely 
explain, for example, why the moderate-luminosity, \hbox{X-ray} selected 
AGN in Figure~10 have notably higher X-ray-to-optical flux ratios 
than the luminous, optically selected quasars. 

Two recent \hbox{X-ray} spectral fitting results on $z>4$ AGN are shown
in Figure~11. Figure~11a shows the \hbox{X-ray} spectrum of the highest
redshift AGN discovered thus far in the \hbox{CDF-N}, a low-luminosity 
quasar at $z=5.186$. It was only possible to obtain a 
respectable-quality \hbox{X-ray} spectrum for such an object due to the 
2~Ms \hbox{CDF-N} exposure. Spectral fitting yields a power-law photon
index of $\Gamma=1.8\pm 0.3$, consistent with observations of
similar objects at low redshift, and there is no evidence for
intrinsic \hbox{X-ray} absorption. Figure~11b shows a ``stacked'' 
spectrum of 46 luminous radio-quiet quasars at $z=4.0$--6.3 (their 
median redshift is $z=4.43$); this spectrum has 750 counts in
total. Joint fitting of the 46 individual spectra, using the 
Cash statistic, yields a power-law photon index ($\Gamma=1.9\pm 0.1$) 
that is again consistent with observations at low redshift. A 
fairly tight limit on any intrinsic \hbox{X-ray} absorption of 
$N_{\rm H}\simlt 9\times 10^{20}$~cm$^{-2}$ is also set. 
The overall picture emerging, then, is that while the AGN 
population shows enormous changes in number density over cosmic 
time, individual AGN \hbox{X-ray} emission regions appear to be
remarkably stable entities. 

% --------------------------------------------------------------------------------

\section{Some Future Prospects}
\label{sec:4}

\begin{figure}[t!]
\centering
\includegraphics[height=10cm]{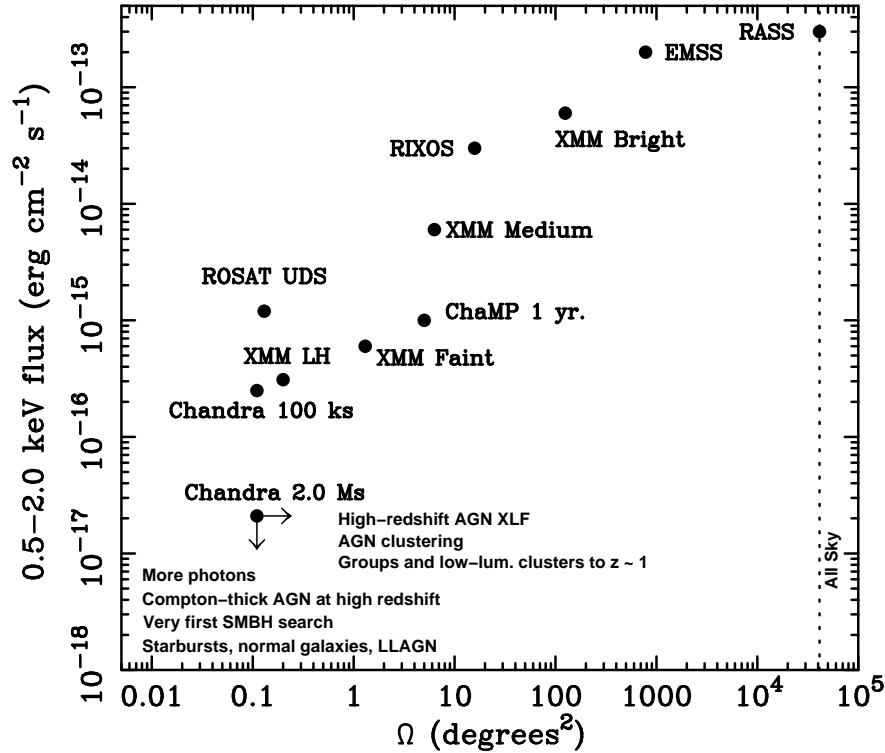}
\caption{A selection of extragalactic X-ray surveys in the \hbox{0.5--2~keV} 
flux limit versus solid angle, $\Omega$, plane. Shown are 
the \rosat\ All-Sky Survey (RASS), 
the \einstein\ Extended Medium-Sensitivity Survey (EMSS), 
the \rosat\ International X-ray/Optical Survey (RIXOS), 
the \xmm\ Serendipitous Surveys ({\it XMM\/} Bright, {\it XMM\/} Medium, {\it XMM\/} Faint), 
the \chandra\ Multiwavelength Project (ChaMP),
the \rosat\ Ultra Deep Survey (\rosat\ UDS), 
the $\approx 100$~ks \xmm\ survey of the Lockman Hole ({\it XMM\/} LH), 
\chandra\ 100~ks surveys, and 
\chandra\ 2~Ms surveys (i.e., the \hbox{CDF-N}). 
Although each of the surveys shown clearly has a range of flux limits 
across its solid angle, we have generally shown the most sensitive 
flux limit. The vertical dot-dashed line shows the solid angle of the 
whole sky. Some key science goals achievable by extending deep \chandra\ 
surveys both wider and deeper are also listed.}
\label{fig:12}       
\end{figure}

\subsection{Future Prospects for Chandra and XMM-Newton}
\label{sec:4p1}

Future prospects for learning more about AGN physics and evolution via \hbox{X-ray}
surveys appear wonderful! Follow-up studies for most of the $\approx 40$ 
surveys listed in Table~1 and Table~2 are ongoing, and many exciting results 
are thus guaranteed even if no more \hbox{X-ray} data are taken. Fortunately, 
however, both \chandra\ and \xmm\ continue to generate torrents of superb 
new data that can provide even more impressive advances. 

\begin{figure}[t!]
\centering
\includegraphics[height=5.7cm]{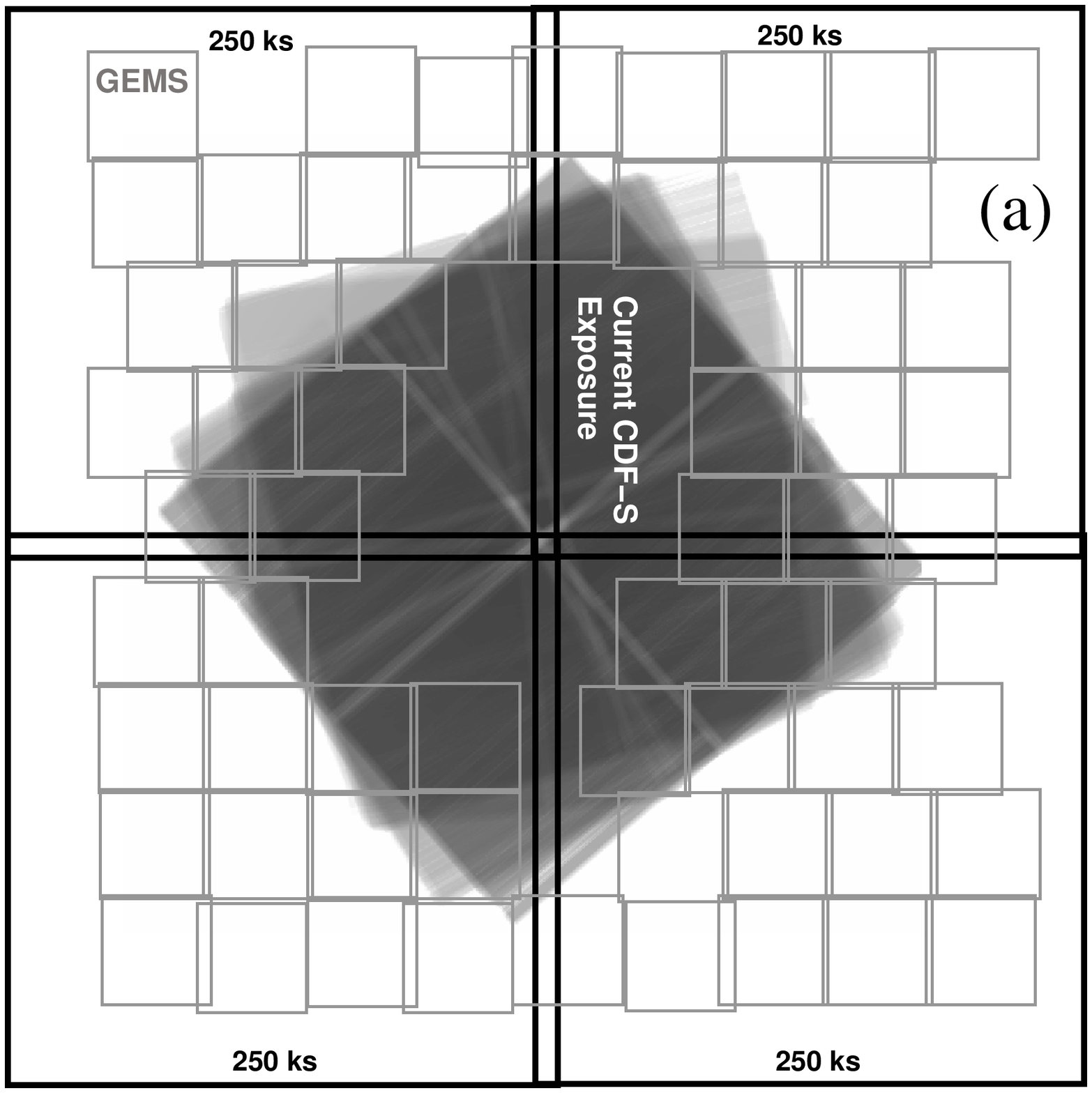}
\includegraphics[height=5.7cm]{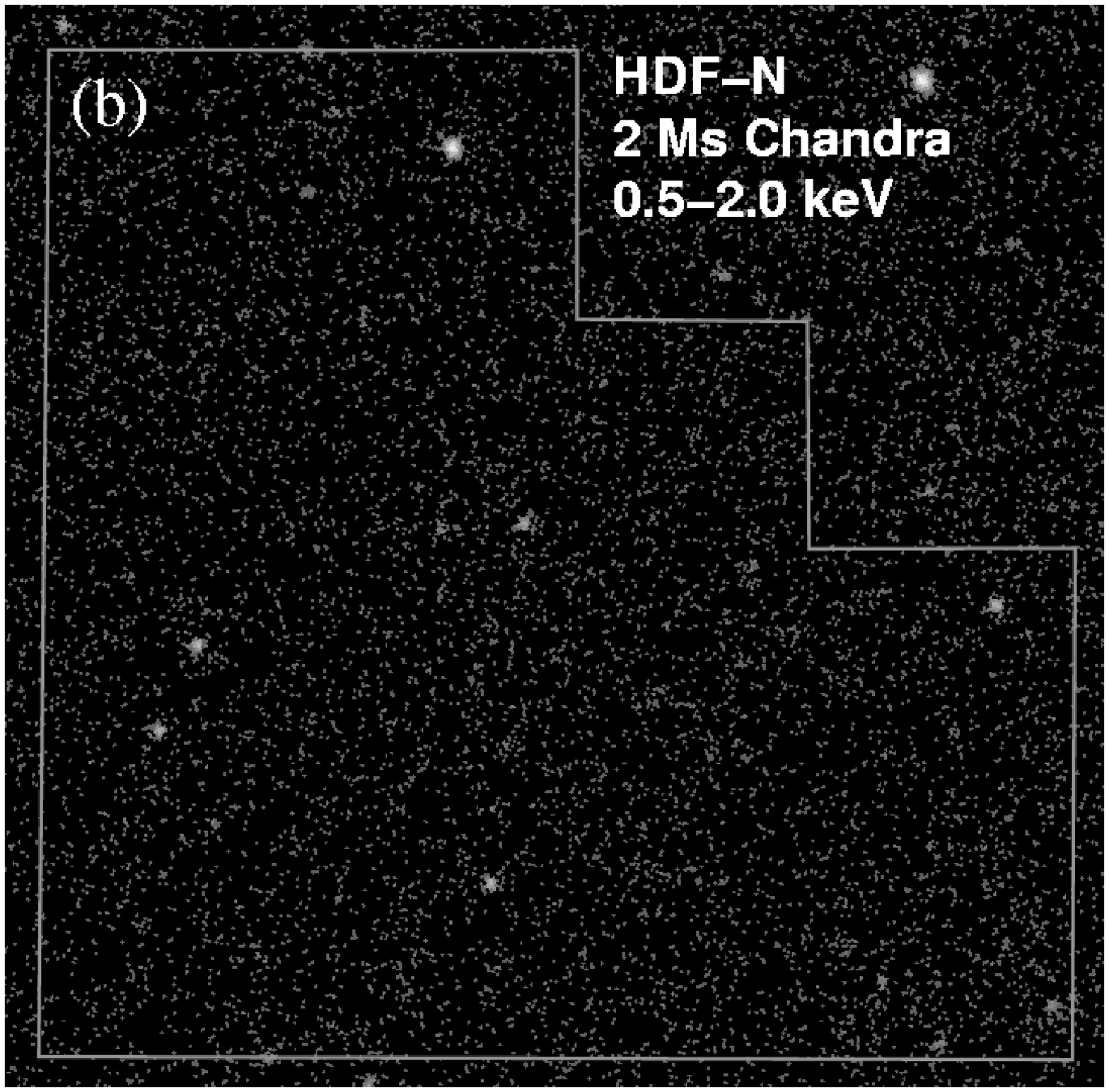}
\caption{(a) Schematic illustration of the Extended \chandra\ Deep 
Field-South survey. The underlying grayscale image shows the current
\hbox{CDF-S} exposure map. The four large black squares show the coverage 
of the upcoming four 250~ks \chandra\ observations. The 63 small 
gray squares show the coverage of \hst\ ACS observations made by the 
GEMS project (the GOODS survey provides \hst\ ACS coverage for the
central region not covered by GEMS).  
(b) \chandra\ \hbox{0.5--2~keV} image of the central part of the 2~Ms \hbox{CDF-N} 
centered on the HDF-N (shown in outline). Note that most ($\approx 94$\%) 
pixels are black, indicating no background. \chandra\ is essentially in the 
photon-limited regime with a 2~Ms exposure, and it can remain in
this regime even with an $\approx 10$~Ms exposure (for 0.5--2~keV sources
near the field center).}
\label{fig:13}       
\end{figure}

Where can the capabilities of \chandra\ and \xmm\ be best applied in 
future observations? Figure~12 presents one useful way of thinking about 
this issue, via a plot of \hbox{0.5--2~keV} flux limit versus solid angle for
selected \hbox{X-ray} surveys. Key parts of this diagram remain to be explored. 
For example, very little solid angle has been surveyed at \hbox{0.5--2~keV} 
flux levels of \hbox{(2--20)$\times 10^{-17}$~erg~cm$^{-2}$~s$^{-1}$}, and thus 
our understanding of the \hbox{X-ray} universe at these flux levels suffers from
limited source statistics and likely cosmic variance. These 
flux levels are below the \xmm\ confusion limit, and thus multiple 
\hbox{0.25--2~Ms} \chandra\ observations are required. Specific science goals 
that can be advanced with this approach include 
(1) pinning down the \hbox{X-ray} luminosity function of moderate-luminosity
AGN at \hbox{$z\approx 2$--6}, 
(2) tracing AGN clustering out to high redshift; this is ideally
done with contiguous, deep coverage, and 
(3) measuring the evolution and properties of groups and low-luminosity
clusters out to $z\approx 1$. 
Figure 13a depicts the ongoing Extended \chandra\ Deep Field-South
survey, which has been guided by the philosophy above. It will
cover a contiguous $\sim 1/4$~deg$^2$ area at a \hbox{0.5--2~keV} flux level 
of \hbox{(1--2)$\times 10^{-16}$~erg~cm$^{-2}$~s$^{-1}$}, and it should
generate $\approx 400$ new AGN (in addition to the $\approx 300$ already
known in the \hbox{CDF-S}). Almost all of these will have superb \hst\ imaging
and multiwavelength coverage. 

An equally important guiding philosophy is to observe one field with \chandra\ 
as sensitively as possible (see Figure~12). Reaching \hbox{0.5--2~keV} flux 
levels of $\approx 5\times 10^{-18}$~erg~cm$^{-2}$~s$^{-1}$ 
is entirely feasible; \chandra\ could remain nearly
photon limited near the field center (see Figure~13b), and source confusion is 
unlikely even for source densities exceeding 100,000~deg$^{-2}$. 
The total required exposure time on a field is $\approx 10$~Ms. Specific
science goals include (1) determining if there is a significant population 
of Compton-thick AGN at \hbox{$z\approx 0.5$--4} that has been missed to date 
(see \S3.1), (2) tightening constraints on moderate-luminosity AGN at 
\hbox{$z\approx 4$--10}, (3) detecting hundreds of normal and starburst galaxies 
out to high redshift (these should outnumber the AGN), and 
using their \hbox{X-ray} emission as an independent, 
extinction-free measure of star-formation rate, and (4) obtaining 
significant numbers of \hbox{X-ray} photons on the faint \hbox{X-ray} source populations 
currently known, so that \hbox{X-ray} spectral and variability analyses can be
applied effectively to determine their nature. Such a sensitive \hbox{X-ray} 
observation will not be possible again for \hbox{10--20} years (see Figure~14)!  
Performing such an observation now can provide information on the sources 
that will be the primary targets of future missions such as \xeus\ 
and \hbox{\it Generation-X\/}; it will thereby bolster the science cases for these missions
and aid their optimal design. 

\subsection{Upcoming and Planned X-ray Missions}
\label{sec:4p2}

\begin{figure}[t!]
\centering
\includegraphics[height=10cm]{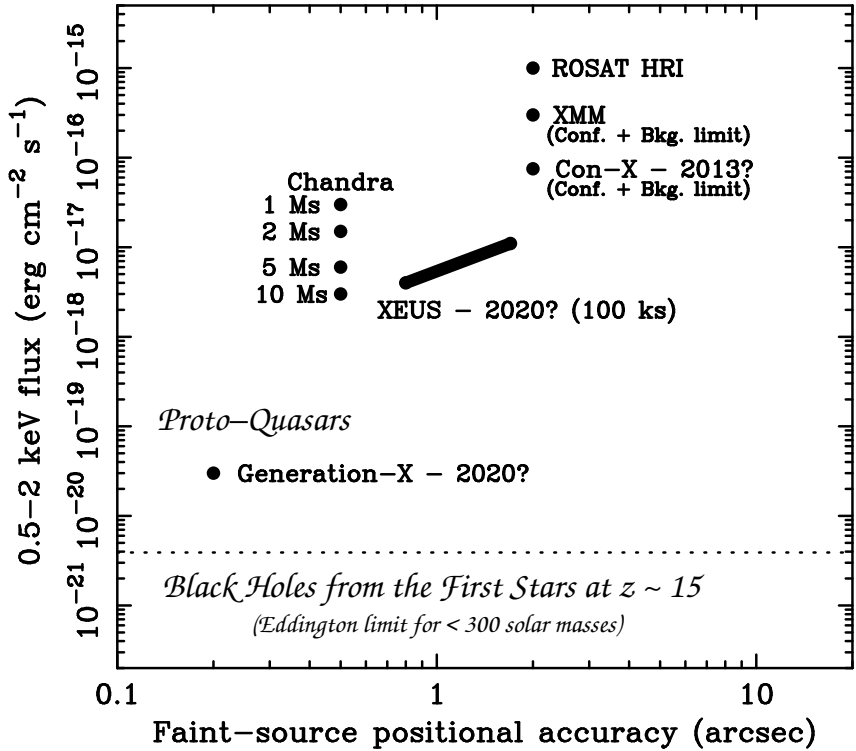}
\caption{Flux limit from \hbox{0.5--2~keV} versus faint-source positional accuracy 
for some past, present, and future \hbox{X-ray} missions (the locations in the diagram 
and launch dates for future missions are approximate). With a \hbox{5--10~Ms} exposure, 
\chandra\ can achieve sensitivities comparable to those discussed for \xeus. 
Furthermore, \chandra\ positions are likely to be the best available for 
$\simgt 15$~yr. Also shown are the expected \hbox{X-ray} fluxes from (1) the black 
holes made by the deaths of the first stars at $z\sim 15$, and (2) proto-quasars 
containing black holes of mass \hbox{$\sim 10^3$--$10^4$~M$_\odot$} at 
\hbox{$z\sim 10$--15}.}
\label{fig:14}       
\end{figure}

In the future, both large ($\simgt US\,\$1$~billion, or 
$\simgt 600$~billion Chilean pesos; see Figure~14) and small-to-medium class 
($\approx US\,\$120$--180~million) \hbox{X-ray} missions should substantially advance 
the AGN \hbox{X-ray} survey work described above. \conx, for example, should 
enable high-quality \hbox{X-ray} spectroscopy for some of the remarkable brighter 
sources found in \hbox{X-ray} surveys. \xeus\ should be able to generate hundreds 
of fields that are as sensitive as the deepest \chandra\ surveys, while
also providing superior photon statistics to those available presently. 
Fitting of high-quality \xeus\ spectra should allow direct redshift determination 
in many cases. Ultimately, \genx\ will 
reach flux limits $\sim 100$ times better than 
those of \chandra\ and \xeus\ (see Figure~14). This improved sensitivity should 
allow detection and study of $\sim 1000$~M$_\odot$ ``proto-quasars'' at 
\hbox{$z\approx 10$--15}, enabling investigation of how the stellar-mass black holes 
made by the deaths of the first stars grew to make the first AGN.

Future small-to-medium class \hbox{X-ray} missions, at least one to be launched 
soon, will sensitively survey large areas of sky at high \hbox{X-ray} energies; 
some will access the poorly explored \hbox{$\approx 10$--200~keV} band covering 
the peak of the XRB. After its \hbox{2004--2005} launch, for example, \swift\ 
will serendipitously conduct the most sensitive \hbox{$\approx 10$--150~keV} 
survey to date with its Burst Alert Telescope. A large fraction of the 
sky should be covered over the lifetime of \swift, and \hbox{$\approx 200$--400}
AGN should be detected. In the \hbox{2007--2010} timeframe, proposed missions
such as the {\it Dark Universe Observatory (DUO)\/} and the 
{\it Nuclear Spectroscopic Telescope Array (NuSTAR)\/} will also 
hopefully conduct sensitive surveys in the \hbox{0.3--8~keV} and 
\hbox{6--80~keV} bands, respectively. {\it DUO\/} would detect $\sim 160,000$ 
AGN in its surveys of the North Galactic Cap (the SDSS area) and South 
Galactic Pole, while {\it NuSTAR\/} would carry the first highly 
sensitive, focusing telescope for $>10$~keV \hbox{X-rays}. 
Other planned small-to-medium class missions include
Japan's {\it Monitor of All-sky X-ray Image (MAXI)\/} and 
{\it New X-ray Telescope (NeXT)\/} 
as well as Europe's {\it LOBSTER\/} and {\it ROSITA\/}.
The {\it Black Hole Finder Probe\/}, defined as part of NASA's Beyond
Einstein program, should ultimately obtain an all-sky census of 
accreting black holes using a wide-field imaging telescope in the 
\hbox{$\approx 10$--600~keV} band. 

% --------------------------------------------------------------------------------

\section*{Acknowledgments}
\label{sec:5}

We gratefully acknowledge support from 
NSF CAREER award AST-9983783,  
CXC grant GO2-3187A,  
the Royal Society (DMA), 
the PPARC (FEB), and
Italian Space Agency contract ASI/I/R/057/02 (CV). 
We thank all of our collaborators. 

% --------------------------------------------------------------------------------

\section*{Some Key Recent References}
\label{sec:6}

\noindent
We have noted the most relevant section above at the end of each reference.
Some references are relevant to more than one section; in this case, we have 
noted the first relevant section. 
Clearly the relevant literature is extensive, and the reader is urged to
consult not only the references below but also the works cited in 
these references. 

\vspace*{0.05 in}

\vspace*{0.05 in}
\noindent
D.M. Alexander, W.N. Brandt, A.E. Hornschemeier, et~al., 2001, AJ, 122, 2156 (\S3.1)

\vspace*{0.05 in}
\noindent
D.M. Alexander, H. Aussel, F.E. Bauer, et~al., 2002, ApJ, 568, L85 (\S3.1)

\vspace*{0.05 in}
\noindent
D.M. Alexander, F.E. Bauer, W.N. Brandt, et~al., 2003, AJ, 125, 383 (\S3.3)

\vspace*{0.05 in}
\noindent
D.M. Alexander, F.E. Bauer, W.N. Brandt, et~al., 2003, AJ, 126, 539 (\S2.1)

\vspace*{0.05 in}
\noindent
D.M. Alexander, F.E. Bauer, S.C. Chapman, et~al., 2004, in 
Multiwavelength Mapping of Galaxy Formation and Evolution, 
ed. R. Bender, A. Renzini (Springer-Verlag, Berlin), in press
(astro-ph/0401129; \S3.3)

\vspace*{0.05 in}
\noindent
O. Almaini, S.E. Scott, J.S. Dunlop, et~al., 2003, MNRAS, 338, 303 (\S3.3)

\vspace*{0.05 in}
\noindent
A.J. Barger, L.L. Cowie, A.T. Steffen, et~al., 2001, ApJ, 560, L23 (\S3.3)

\vspace*{0.05 in}
\noindent
A.J. Barger, L.L. Cowie, P. Capak, et~al., 2003, AJ, 126, 632 (\S3.1)

\vspace*{0.05 in}
\noindent
A.J. Barger, L.L. Cowie, P. Capak, et~al., 2003, ApJ, 584, L61 (\S3.4)

\vspace*{0.05 in}
\noindent
F.E. Bauer, D.M. Alexander, W.N. Brandt, et~al., 2002, AJ, 124, 2351 (\S3.1)

\vspace*{0.05 in}
\noindent
A.W. Blain, I. Smail, R.J. Ivison, et~al., 2002, Phys. Rep., 369, 111 (\S3.3)

\vspace*{0.05 in}
\noindent
W.N. Brandt, A. Laor, B.J. Wills, 2000, ApJ, 528, 637 (\S3.1)

\vspace*{0.05 in}
\noindent
W.N. Brandt, A.E. Hornschemeier, D.M. Alexander, et~al., 2001, AJ, 122, 1 (\S3.1)

\vspace*{0.05 in}
\noindent
W.N. Brandt, D.M. Alexander, A.E. Hornschemeier, et~al., 2001, AJ, 122, 2810 (\S2.1)

\vspace*{0.05 in}
\noindent
W.N. Brandt, C. Vignali, D.P. Schneider, et~al., 2004, Adv. Space Res., 
in press (astro-ph/0212082; \S3.4)

\vspace*{0.05 in}
\noindent
S.C. Chapman, A.W. Blain, R.J. Ivison, et~al., 2003, Nature, 422, 695 (\S3.3)

\vspace*{0.05 in}
\noindent
A. Comastri, M. Mignoli, P. Ciliegi, et~al., 2002, ApJ, 571, 771 (\S3.1) 

\vspace*{0.05 in}
\noindent
L.L. Cowie, A.J. Barger, M.W. Bautz, et~al., 2003, ApJ, 584, L57 (\S3.2)

\vspace*{0.05 in}
\noindent
S. Cristiani, D.M. Alexander, F.E. Bauer, et~al., 2004, ApJ, 600, L119 (\S3.4)

\newpage

\vspace*{0.05 in}
\noindent
S.M. Croom, R.J. Smith, B.J. Boyle, et~al., 2004, MNRAS, in 
press (astro-ph/0403040; \S3.2) 

\vspace*{0.05 in}
\noindent
F. Fiore, M. Brusa, F. Cocchia, et~al., 2003, A\&A, 409, 79 (\S3.2)

\vspace*{0.05 in}
\noindent
P. Gandhi, C.S. Crawford, A.C. Fabian, et~al., 2004, MNRAS, 348, 529 (\S3.1)

\vspace*{0.05 in}
\noindent
R. Giacconi, A. Zirm, J.X. Wang, et~al., 2002, ApJS, 139, 369 (\S2.1)

\vspace*{0.05 in}
\noindent
M. Giavalisco, H.C. Ferguson, A.M. Koekemoer, et~al., 2004, ApJ, 600, L93 (\S2.1)

\vspace*{0.05 in}
\noindent
R. Gilli, 2004, Adv. Space Res., in press (astro-ph/0303115; \S3.1)

\vspace*{0.05 in}
\noindent
P.J. Green, J.D. Silverman, R.A. Cameron, et~al., 2004, ApJS, 150, 43 (\S2.2)

\vspace*{0.05 in}
\noindent
G. Hasinger, B. Altieri, M. Arnaud, et~al., 2001, A\&A, 365, L45 (\S2.1)

\vspace*{0.05 in}
\noindent
A.E. Hornschemeier, W.N. Brandt, G.P. Garmire, et~al., 2000, ApJ, 541, 49 (\S3.3)

\vspace*{0.05 in}
\noindent
A.E. Hornschemeier, F.E. Bauer, D.M. Alexander, et~al., 2003, AJ, 126, 575 (\S3.1) 

\vspace*{0.05 in}
\noindent
A.E. Hornschemeier, D.M. Alexander, F.E. Bauer, et~al., 2004, ApJ, 600, L147 (\S3.1)

\vspace*{0.05 in}
\noindent
M.P. Hunt, C.C. Steidel, K.L. Adelberger, et~al., 2004, ApJ, in 
press (astro-ph/0312041; \S3.1)

\vspace*{0.05 in}
\noindent
F. Jansen, D. Lumb, B. Altieri, et~al., 2001, A\&A, 365, L1 (\S1)

\vspace*{0.05 in}
\noindent
D.W. Kim, R.A. Cameron, J.J. Drake, et~al., 2004, ApJS, 150, 19 (\S2.2)

\vspace*{0.05 in}
\noindent
A.M. Koekemoer, D.M. Alexander, F.E. Bauer, et~al., 2004, ApJ, 600, L123 (\S3.4)

\vspace*{0.05 in}
\noindent
A. Marconi, G. Risaliti, R. Gilli, et~al., 2004, MNRAS, 
in press (astro-ph/0311619; \S3.1)

\vspace*{0.05 in}
\noindent
G. Matt, 2002, Phil. Trans. R. Soc. Lond. A, 360, 2045 (\S3.1)

\vspace*{0.05 in}
\noindent
G. Matt, A.C. Fabian, M. Guainazzi, et~al., 2000, MNRAS, 318, 173 (\S3.1) 

\vspace*{0.05 in}
\noindent
E.C. Moran, A.V. Filippenko, R. Chornock, 2002, ApJ, 579, L71 (\S3.1)

\vspace*{0.05 in}
\noindent
R.F. Mushotzky, L.L. Cowie, A.J. Barger, et~al., 2000, Nature, 404, 459 (\S2.1)

\vspace*{0.05 in}
\noindent
B.M. Peterson, 1997, An Introduction to Active Galactic Nuclei (Cambridge
University Press, Cambridge), pp. 183--193 (\S3.2)

\vspace*{0.05 in}
\noindent
G. Risaliti, R. Maiolino, M. Salvati, 1999, ApJ, 522, 157 (\S3.1)

\vspace*{0.05 in}
\noindent
H.-W., Rix, M. Barden, S.V.W. Beckwith, et~al., 2004, ApJ, 
in press (astro-ph/0401427; \S4.1)

\vspace*{0.05 in}
\noindent
M. Schmidt, 1968, ApJ, 151, 393 (\S3.2)

\vspace*{0.05 in}
\noindent
D.P. Schneider, X. Fan, P.B. Hall, et~al., 2003, AJ, 126, 2579 (\S3.4)

\vspace*{0.05 in}
\noindent
P. Severgnini, A. Caccianiga, V. Braito, et~al., 2003, A\&A, 406, 483 (\S3.1)

\vspace*{0.05 in}
\noindent
I.A.G. Snellen, P.N. Best, 2001, MNRAS, 328, 897 (\S3.1)

\vspace*{0.05 in}
\noindent
A.T. Steffen, A.J. Barger, L.L. Cowie, et~al., 2003, ApJ, 596, L23 (\S3.2) 

\vspace*{0.05 in}
\noindent
G.P. Szokoly, J. Bergeron, G. Hasinger, et~al., 2004, ApJS, 
submitted (astro-ph/0312324; \S3.1)

\newpage

\vspace*{0.05 in}
\noindent
Y. Ueda, M. Akiyama, K. Ohta, et~al., 2003, ApJ, 598, 886 (\S3.1)

\vspace*{0.05 in}
\noindent
C. Vignali, F.E. Bauer, D.M. Alexander, et~al., 2002, ApJ, 580, L105 (\S3.4)

\vspace*{0.05 in}
\noindent
C. Vignali, W.N. Brandt, D.P. Schneider, 2003, AJ, 125, 433 (\S3.4)

\vspace*{0.05 in}
\noindent
C. Vignali, W.N. Brandt, D.P. Schneider, et~al., 2003, AJ, 125, 2876 (\S3.4)

\vspace*{0.05 in}
\noindent
M.C. Weisskopf, H.D. Tananbaum, L.P. Van Speybroeck, et~al., 2000, 
Proc. SPIE, 4012, 2 (\S1)

\vspace*{0.05 in}
\noindent
R.E. Williams, B. Blacker, M. Dickinson, et~al., 1996, AJ, 112, 1335 (\S2.1)

\vspace*{0.05 in}
\noindent
C. Wolf, L. Wisotzki, A. Borch, et~al., 2003, A\&A, 408, 499 (\S3.2)

\vspace*{0.05 in}
\noindent
D.G. York, J. Adelman, J.E. Anderson, et~al., 2000, AJ, 120, 1579 (\S3.4)

\vspace*{0.05 in}
\noindent
F. Yuan \& R. Narayan, 2004, ApJ, in press (astro-ph/0401117; \S3.1)

% \printindex

\end{document}